\documentclass[]{jfm}
\usepackage{graphicx}
\usepackage{newtxtext}
\usepackage{newtxmath}
\usepackage{natbib}
\usepackage{hyperref}
\hypersetup{
    colorlinks = true,
    urlcolor   = blue,
    citecolor  = black,
}
\usepackage[]{ar}

\newcommand{\RomanNumeralCaps}[1]

\shorttitle{Wall-attached and VLSM streamwise velocity energy in the atmospheric surface layer}
\shortauthor{Puccioni et al.}

\title{Identification of the energy contributions associated with wall-attached eddies and very-large-scale motions in the near-neutral atmospheric surface layer through wind LiDAR measurements}

\author{Matteo Puccioni\aff{1},
    Marc Calaf\aff{2},
    Eric R. Pardyjak\aff{2},
    Sebastian Hoch\aff{2},
    Travis J. Morrison\aff{2},
    Alexei Perelet\aff{2},
  Giacomo Valerio Iungo\aff{1}\corresp{\email{valerio.iungo@utdallas.edu}}}

\affiliation{\aff{1}Wind Fluids and Experiments (WindFluX) Laboratory, Mechanical Engineering Department, The University of Texas at Dallas, 800 W Campbell Rd, 75080 Richardson, Texas, USA{\aff{2}University of Utah}}

\begin{document}
\maketitle

\begin{abstract}
Recent works on wall-bounded flows have corroborated the coexistence of wall-attached eddies, whose statistical features are predicted through Townsend's attached eddy hypothesis (AEH), and very-large-scale motions (VLSMs), which are not encompassed in the AEH. Furthermore, it has been shown that the presence of wall-attached eddies within the logarithmic layer is linked to the appearance of an inverse-power-law region in the streamwise velocity energy spectra, upon significant separation between outer and viscous scales. In this work, a near-neutral atmospheric surface layer (ASL) is probed with a wind LiDAR to investigate the contributions to the streamwise velocity energy associated with wall-attached and VLSMs for a very-high Reynolds-number boundary layer. Energy and linear coherence spectra (LCS) of the streamwise velocity are interrogated to identify the spectral boundaries associated with eddies of different typologies and the maximum height attained by wall-attached eddies. Inspired by the AEH, an analytical model for the LCS associated with wall-attached eddies is formulated. The experimental results show that the identification of the wall-attached-eddy energy contribution through the analysis of the energy spectra leads to an underestimate of the associated spectral range, maximum height attained, and turbulence intensity. This feature is due to the overlap of the energy associated with VLSMs obscuring the inverse-power-law region. On the other hand, the Townsend-Perry constant for the turbulence intensity seems to be properly estimated through the spectral analysis. The LCS analysis estimates wall-attached eddies with a streamwise/wall-normal ratio of about 14.3 attaining a height of about 30\% of the outer scale of turbulence.
\end{abstract}


\section{Introduction}\label{sec: Intro}
Characterizing the organization and energy content of coherent structures present in wall-bounded turbulent flows is important for many engineering and environmental pursuits, such as wind energy \citep{Onder2018}, environmental pollutant transport \citep{Reche2018}, and urban flows \citep{Barlow2014}. Coherent structures cover a breadth of spatial and temporal scales, such as streamwise-aligned packets of hairpin vortices, named large-scale motions (LSMs), characterized by streamwise-elongated velocity fluctuations with wavelengths comparable to the outer scale of turbulence, $\Delta_E$ (e.g., the boundary layer thickness) \citep{Adrian2007,Marusic2010,Smits2011,Jimenez2018}. For very high Reynolds numbers, coherent structures even longer than $\Delta_E$, namely very-large-scale motions (VLSMs) \citep{Kim1999,Guala2006,Balakumar2007,Hutchins2007a} are thought to arise from streamwise aggregations of LSMs \citep{Kim1999}, and interact with the near-wall turbulence cycle through non-linear modulation mechanisms \citep{Mathis2009,Talluru2014,Liu2019,Lee2019,Salesky2020}.

A cornerstone to achieving an in-depth understanding of the stochastic contribution of coherent structures to the turbulent kinetic energy in wall-bounded flows is the Townsend's attached-eddy hypothesis (AEH) \citep{Townsend1976}, which models the logarithmic layer as a forest of randomly-repeated geometrically-similar eddies, whose vertical extent, $\delta$, is proportional to their distance from the wall, $z$, and whose eddy population density is inversely proportional to their size. Furthermore, the geometric similarity of wall-attached turbulent motions and the overlapping between inner-scaling with $z$ and outer-scaling with $\Delta_E$ justify the presence of the $k_x^{-1}$ region (where $k_x$ is the streamwise wavenumber) in the streamwise velocity energy spectrum, $\phi_{uu}$ \citep{Perry1986}. This spectral feature was also predicted through dimensional analysis \citep{Perry1975,Perry1986,Davidson2009}. The spectral extension of this inverse-power-law region is expected to grow with scale separation, and, thus, the friction Reynolds number $\Rey_\tau=U_\tau\Delta_E/\nu$, where $U_\tau$ is the friction velocity ($U_\tau = \sqrt{\tau_0/\rho}$, with $\tau_0$ and $\rho$ being the wall-shear stress and the fluid density, respectively), and $\nu$ the kinematic viscosity. However, evidence of the $k_x^{-1}$ spectral region is still elusive even for high $\Rey_\tau$ laboratory data, \citep{Morrison2002,Rosenberg2013,Vallikivi2015,Baidya2017}, and field observations of the atmospheric surface layer (ASL) \citep{Hogstrom2002,Calaf2013}. 

\cite{Perry1977,Perry1986,Marusic1995,Perry1995} argued that the coherent structures in wall-bounded flows do not consist of only wall-attached eddies, rather they encompass eddies of different nature. In this scenario, \cite{Perry1995} reasoned that three different eddy types exist in a wall-bounded flow: wall-attached eddies described by the AEH (type $\mathcal{A}$ eddies), wall-detached eddies (type $\mathcal{B}$ eddies), referring to large-scale structures, superstructures and VLSMs \citep{Hogstrom1990,Hogstrom1992,Hogstrom2002,Baars2020a,Hu2020}, and Kolmogorov small-scale eddies (type $\mathcal{C}$), which dominate the $k_x^{-5/3}$ inertial sub-range of the streamwise velocity spectrum. Despite the capability of the AEH in providing an accurate representation of the stochastic energetic contribution of wall-attached eddies to the logarithmic layer of wall-bounded turbulent flows, the stochastic identification of turbulent coherent structures of different nature, i.e. type $\mathcal{A}$, $\mathcal{B}$, or $\mathcal{C}$ eddies according to the classification proposed by \cite{Perry1995}, is still elusive. 

A common technique to separate the energetic contributions due to coherent structures and, specifically, to isolate the energy connected with wall-attached eddies, is to apply a band-pass filter to the streamwise velocity signals \citep[e.g.][]{Nickels2005,Hwang2015,Hu2020}. The AEH assumes that wall-attached eddies scale as their wall-normal distance, and, thus, the high-wavenumber limit of a band-pass filter aiming at isolating the wall-attached-eddy contribution from the streamwise velocity spectrum should be proportional to $z$ \citep{Perry1982,Meneveau2013,Yang2019,Hu2020,Baars2020a}. Furthermore, the streamwise velocity within the logarithmic layer at a given wall-normal position results from the superposition of contributions induced by wall-attached eddies within the vertical range between $z$ and $\Delta_E$. Therefore, the low-frequency limit of a potential band-pass filter should scale with the boundary layer thickness, $\Delta_E$ \citep{Baars2020a,Hu2020}. While there is consensus on the filtering approach to isolate the energetic contribution associated with wall-attached eddies, on the other hand, there are broad discrepancies on the actual spectral limits used for this band-pass filter. 

Another technique to separate the energy content associated with coherent structures of different nature was proposed in \cite{Baars2020a}. In their study, the authors generated two spectral filters based on the linear coherence spectrum (LCS) obtained from the streamwise-velocity signals collected at a given height and two reference positions, one located in the proximity of the wall and another within the logarithmic layer. Using this data-driven approach, the authors found that the coherence-based low-wavelength limit of the $k_x^{-1}$ spectral region was proportional to the wall-normal position ($\lambda_x\geq14~z$, where $\lambda_x$ is the streamwise wavelength), while the high-wavelength limit was proportional to $\Delta_E$. 

As mentioned above, the spectral extension of the inverse-power-law region grows with the separation between the outer scale of turbulence, $\Delta_E$, and the viscous scale, $\nu/U_\tau$. This requirement has spurred the development of experimental facilities \citep{Marusic2010,Smits2011,Marusic2018} and numerical tools \citep{Jimenez2004,Jimenez2007,Lee2015,Lee2019} enabling investigations of wall-bounded flows at high Reynolds numbers. For the same purpose, the ASL represents a unique opportunity to probe a boundary layer with extremely-high Reynolds numbers \citep{Metzger2007,Guala2011,Hutchins2012,Liu2017a,Heisel2018,Li2021,Huang2021} upon filtering out velocity fluctuations connected with non-turbulent scales \citep{Larsen2013,Larsen2016}, restricting the data set to subsets presenting negligible effects connected with the atmospheric thermal stratification \citep{Wilson2008}, and strictly verifying the statistical stationarity and convergence of the collected measurements \citep{Metzger2007}. Analogies between ASL and laboratory flows have already been investigated for several features of turbulent boundary layers, such as near-wall structures \citep{Klewicki1995}, hairpin vortex packets \citep{Hommema2003}, Reynolds stresses \citep{Kunkel2006,Marusic2013}, inclination angle of coherent structures \citep{Liu2017a}, uniform momentum zones \citep{Heisel2018}, large-scale amplitude modulation process \citep{Liu2019}, and LCS \citep{Krug2019,Li2021}. 

Probing ASL flows requires measurement techniques providing sufficient spatio-temporal resolution throughout the ASL thickness. In this realm, wind light detection and ranging (LiDAR) has become a compelling remote sensing technique to investigate atmospheric turbulence. For instance, LiDAR scans can be optimally designed to probe the atmospheric boundary layer and wakes generated by utility-scale wind turbines \citep[e.g]{Letizia2021a,ElAsha2017,Zhan2020,Letizia2021b}. Regarding atmospheric turbulence, LiDAR measurements were used to detect the inverse-power law \citep{Calaf2013} or the inertial sub-layer \citep{Iungo2013} from the streamwise velocity energy spectra. Multiple simultaneous and co-located LiDAR measurements can be leveraged to measure 3D velocity components and Reynolds stresses \citep{Mann2009,Mikkelsen2008,Carbajo2014}. More recently, the LiDAR technology was assessed against sonic anemometry during the eXperimental Planetary boundary layer Instrumentation Assessment (XPIA) campaign \citep{Debnath2017a,Lundquist2017,Debnath2017b}.

In this work, streamwise-velocity measurements collected simultaneously at various wall-normal positions throughout the ASL thickness with a pulsed Doppler scanning wind LiDAR and a sonic anemometer are investigated to identify the spectral boundaries and the maximum vertical extent of the energy contributions associated with wall-attached eddies. The velocity data were collected through fixed scans performed with the azimuth angle of the scanning head set along the mean wind direction during near-neutral thermal conditions. After the quantification of the spectral gap and estimation of the outer scale of turbulence, $\Delta_E$, the identification of the energy associated with eddies of different nature is performed through two independent methods: first, from the streamwise velocity energy spectra by leveraging the semi-empirical spectral model proposed by \cite{Hogstrom2002}; then, from the LCS obtained between the LiDAR data collected at a reference height and various wall-normal positions, in analogy with the approach proposed by \cite{Baars2017}. Finally, the integrated streamwise energy within both the spectral portion associated with wall-attached eddies, and that due to coherent structures with larger wavelengths, e.g. VLSMs and superstructures, is evaluated along the wall-normal direction and assessed against previous laboratory studies. 

The remainder of the paper is organized as follows: in \S\ref{sec: data set}, the experimental data set is introduced, while the methodology to analyze the streamwise velocity spectrum and LCS is described in \S\ref{sec: Theory}.  In \S\ref{sec: Results}, the results on the identification of turbulent coherent structures of different nature and the quantification of their energy content are discussed. Finally, concluding remarks are reported in \S\ref{sec: Conclusion}. 

In this work, a Cartesian reference frame is used, where 
$(x,~y,~z)$ represent the streamwise, spanwise and wall-normal directions, respectively. The respective mean velocity vector is indicated as $\textbf{U}=(U,~V,~W)$, while the zero-mean velocity fluctuations are $\textbf{u}=(u,~v,~w)$. Overbar denotes the Reynolds average, $t$ is time, and the superscript “$+"$ for a dimension of length indicates the viscous scaling with $\nu/U_\tau$, while outer scaling is performed via $\Delta_E$.

\section{Experimental data set}\label{sec: data set}
\subsection{The LiDAR field campaign}\label{subsec: LiDARfield}
\begin{figure}
    \centerline{\includegraphics[width=\textwidth]{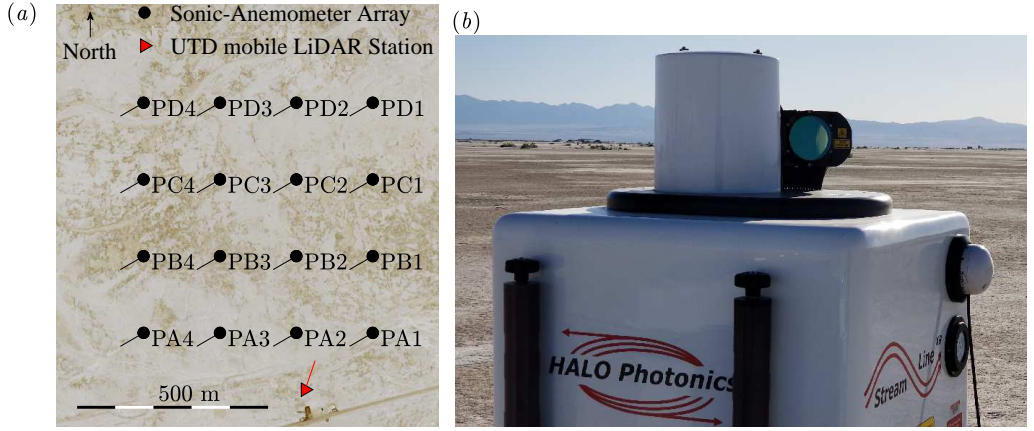}}
    \caption{LiDAR field campaign: (\emph{a}) Aerial view of the instrument locations. Lines represent the orientation of each instrument, while the labels report names of each sonic anemometer; (\emph{b}) Photo of the scanning Doppler wind LiDAR.}\label{fig: IPAQS}
\end{figure}

Wind and atmospheric data were collected during the Idealized Planar Array experiment for Quantifying Surface heterogeneity (IPAQS) campaign performed in June 2018 at the Surface Layer Turbulence and Environmental Science Test (SLTEST) site \citep{Huang2021}. This site is located in the South-West part of the dry Great Salt Lake, Utah, within the Dugway Proving Ground military facility. The SLTEST site is characterized by an extremely barren and flat ground ($\approx$1 m elevation difference every $13$ km) and exceptionally long extensions ($\approx$240 km and $\approx$48 km in the North-South and East-West directions, respectively) \citep{Kunkel2006}. The typical terrain coverage consists of bushes and small hills, which, combined with the dry and salty soil, classify the terrain as transitionally rough \citep{Ligrani1986,Kunkel2006}. During the experimental campaign, several instruments were simultaneously deployed for different scientific purposes. The University of Texas at Dallas (UTD) mobile LiDAR station (red triangle in figure \ref{fig: IPAQS}\emph{a}) was deployed in the proximity of a 4-by-4 array of CSAT3 3D sonic anemometers (black circles in figure \ref{fig: IPAQS}\emph{a}), manufactured by Campbell Scientific Inc., which recorded the three velocity components and temperature with a sampling frequency of 20 Hz. The sonic anemometer data considered for this study were collected from the station indicated as “PA2” in figure \ref{fig: IPAQS}(\emph{a}) at a 2-m height. The sonic-anemometer data were firstly corrected for pitch and yaw misalignment following the procedure proposed by \cite{Wilczak2001}, then high-pass filtered as per \cite{Hu2020} with cut-off frequency $f_\text{gap}=0.0055$ Hz, whose selection is discussed in Appendix \ref{subsec: SpectralGap}.

To investigate a canonical near-neutral boundary layer, the effects of atmospheric stability on the velocity field should be accounted for. Regarding atmospheric boundary layer flows, the buoyancy contribution to turbulence is compared to the shear-generated turbulence through the Obukhov length, $L$ \citep{Monin1954}:
\begin{equation}\label{eq: Obukhov}
    L = -\frac{U_{\tau}^3 T}{\kappa g \overline{w\theta}}~,
\end{equation}
where $T$ is the mean virtual potential temperature (in Kelvin), $\kappa=0.41$ is the von K\'{a}rm\'{a}n constant, $\overline{w\theta}$ is the vertical heat flux, and $g$ is the gravity acceleration. Sonic-anemometer data from the PA2 station are further leveraged to calculate the friction velocity according to the eddy-covariance method \citep{Stull1988}:
\begin{equation}\label{eq: FrictionVelocity}
    U_{\tau} = (\overline{uw}^2 + \overline{vw}^2)^{1/4}.
\end{equation}

The pulsed scanning Doppler wind LiDAR deployed for this experiment is a Streamline XR manufactured by Halo Photonics, whose technical specifications are reported in table \ref{tab:HaloSpec} while a photo of its deployment for the IPAQS field campaign is reported in figure \ref{fig: IPAQS}(\emph{b}). A Doppler wind LiDAR allows probing the atmospheric wind field utilizing a laser beam whose light is backscattered due to the presence of particulates suspended in the atmosphere. The velocity component along the laser-beam direction, denoted as radial velocity, $V_r$, is evaluated from the Doppler shift of the backscattered laser signal \citep{Sathe2013}. A pulsed Doppler wind LiDAR emits laser pulses to perform quasi-simultaneous wind measurements at multiple distances from the LiDAR as the pulses travel in the atmosphere. The wind measurements performed over each probe volume, which is referred to as range gate, can be considered as the convolution of the actual wind velocity field projected along the laser-beam direction with a weighting function representing the radial distribution of the energy associated with each laser pulse \citep{Frehlich1998}. Therefore, the radial velocity, $V_r$, can be expressed in terms of the instantaneous wind velocity components, $(U(t),~V(t),~W(t))$, where the $x$-direction is considered aligned with the mean wind direction, $\Theta_w$, as:
\begin{equation}\label{eq: RadialVelocity}
    V_r(t) = U(t)\cos\Theta\cos\Phi + V(t)\sin\Theta\cos\Phi + W(t)\sin\Phi,
\end{equation}
where $\Theta$ is the LiDAR azimuth angle, and $\Phi$ is the elevation angle.

To maximize the spatio-temporal resolution of the LiDAR measurements and accuracy in probing the streamwise velocity component, fixed LiDAR scans were performed with a low elevation angle ($\Phi=3.5^{\circ}$) and with the laser beam aligned with the mean wind direction, which is monitored by the PA2 sonic anemometer, namely with $V\approx 0$ \citep{Iungo2013}. During the post-processing, only LiDAR data sets with an instantaneous deviation of the  wind direction smaller than $\pm10^{\circ}$ from the respective 10-minute average have been considered \citep{Hutchins2012}. Considering the low elevation angle used and the azimuth angle aligned with the mean wind direction, the first-order approximation for the mean streamwise velocity measured through the wind LiDAR is obtained from (\ref{eq: RadialVelocity}) as $U \approx V_r/\cos{\Phi}$, while for the variance is $\overline{uu} \approx \overline{v_rv_r}/\cos^2{\Phi}$ \citep{Zhan2020}. 

As previously mentioned, the LiDAR radial velocity is measured through a convolution of the LiDAR laser pulse with the actual velocity field over each range gate. This spatial averaging leads to an underestimation of the measured streamwise turbulence intensity. In this work, the streamwise velocity energy spectra, and the respective turbulence intensity obtained as integral over the measured spectral range, are corrected for the spatial averaging associated with the LiDAR measuring process by using the methodology proposed in \cite{Puccioni2021}. The reader can refer to Appendix \ref{sec: SpectraCorrection} for more details.

Based on the data quality control described in the following subsection \S\ref{subsec: QualityControl}, a subset of one hour of LiDAR data collected during the day of June 10, 2018, from 09:00 AM to 10:00 AM UTC (local time -6 hours) is selected for further analyses, which is characterized by a friction velocity of $U_{\tau}=0.42~ \text{m}~\text{s}^{-1}$, Obukhov length of $L=-278$ m, which corresponds to a stability parameter of $z/L=-0.007$ indicating a near-neutral atmospheric stability regime \citep{Hogstrom2002,Kunkel2006,Metzger2007,Mouri2019,Huang2021}. For the selected data set, the kinematic viscosity has been estimated $\nu=1.49\times10^{-5}~ \text{m}^2~\text{s}^{-1}$ based on the mean temperature of $290$ K recorded by the sonic anemometer \citep{Picard2008}.
\begin{table}
\begin{center}
\def~{\hphantom{0}}
\begin{tabular}{lc}
{Parameter}                  &{{Value}}           \\[3pt]
Wavelength ($\mu$m)                & 1.5   \\
Repetition rate (kHz)            & 10  \\
Velocity resolution (m s$^{-1}$)       & $\pm 0.0764$  \\
Velocity bandwidth (m s$^{-1}$) & $\pm 38$ \\
Number of FFT points       & 1024     \\
Radial range (m)         & $45$ to $10000$ \\
Azimuth angle (range) ($^\circ$)      & $0^\circ$ to $360^\circ$\\
Elevation angle (range) ($^\circ$)    & $-10^\circ$ to $190^\circ$\\
LiDAR gate length (m)         & 18   \\
Number of gates            & 200       \\
Sampling rate (Hz)     			& 1   \\
\end{tabular}
\caption{Technical specifications of the scanning Doppler wind LiDAR Streamline XR.\label{tab:HaloSpec}} \end{center}
\end{table}

\subsection{Quality control of the LiDAR data}\label{subsec: QualityControl}
LiDAR measurements undergo a quality control procedure to ensure reliability and accuracy of the velocity data. The first parameter used to ensure the accuracy of the LiDAR velocity measurements is the carrier-to-noise ratio (CNR), which represents a quantification of the intensity of the backscattered laser pulse over the typical signal noise as a function of the radial distance and time \citep{Frehlich1997,Beck2017,Gryning2019}. For a fixed scan with a constant elevation angle, the range gates selected for any further analysis have a time-averaged CNR not lower than $-20$ dB \citep{Gryning2019}, which corresponds for the selected data set to all the LiDAR measurements collected within the vertical range between 6 m and 143 m with a vertical resolution of approximately 1.08 m. Considering an elevation angle of the laser beam of $3.5^\circ$, the horizontal range between the first and last LiDAR range gate is then $2246$ m. 

A filtering  procedure is then adopted to remove possible outliers from the LiDAR data, i.e. erroneous estimation of the radial velocity from the backscattered LiDAR signal \citep{Frehlich1998}. In this study, a standard deviation-based filter is implemented, i.e. any velocity sample out of the interval $U\pm 3.5 {\overline{uu}}^{1/2}$, which is estimated for the entire 1-hour period of the selected data set, is marked as an outlier and removed \citep{Hojstrup1993,Vickers1997}. The rejected samples are then replaced through a bi-harmonic algorithm with the Matlab function $inpaint\_nans$ \citep{DErrico2004}. In the worst scenario, the total number of samples rejected for the data collected at a 140-m height is $0.75\%$ over the 1-hour duration of the selected data set. The LiDAR signal quality typically improves by approaching the LiDAR due to the increased energy in the laser beam.

\begin{figure}
    \centering
   \centerline{\includegraphics{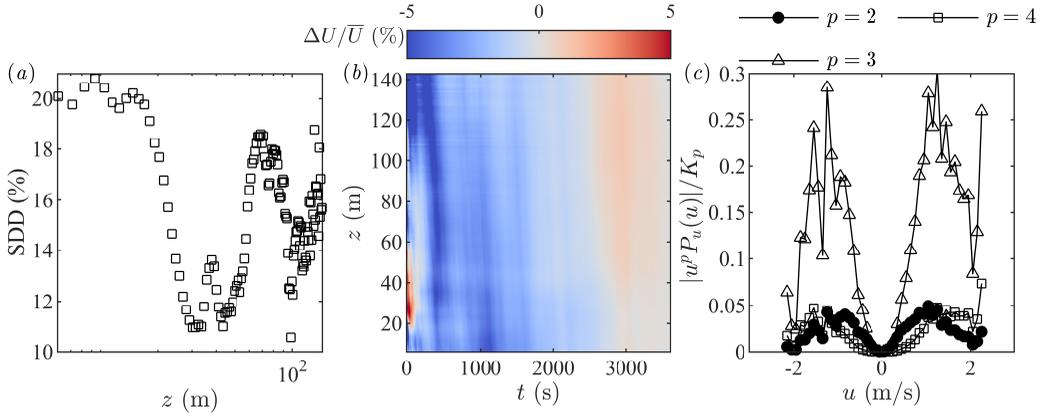}}
    \caption{Analysis of the statistical stationarity and convergence of the LiDAR data: (\textit{a}) $\textrm{SDD}$ parameter as a function of height; (\textit{b}) Percentage difference between cumulative mean for different signal durations, $t$, and the mean for the entire 1-hour duration of the LiDAR velocity data, $\overline{U}$; (\textit{c}) Normalized absolute value of pre-multiplied probability density functions for the velocity signal collected at $z$=143 m for statistical moments with a different order, $p$. \label{fig:StatisticalConvergence}}
\end{figure}

Subsequently, the statistical stationarity of the velocity signals is analyzed through the standard deviation dispersion, $\textrm{SDD}$, which is calculated as follows \citep{Foken1996}:

\begin{equation}\label{eq: SDD}
    \text{SDD}=\dfrac{\sqrt{\langle (\sigma_i-\sigma_0)^2\rangle_i}}{\sigma_0}\times100,
\end{equation}
where $\sigma_0$ is the standard deviation of a velocity signal over its entire sampling period of 1 hour, and $\sigma_i$ is the velocity standard deviation calculated over a subset with a 5-minute duration. The symbol $\langle \rangle_i$ indicates the average over the total number of non-overlapping subsets of the original velocity signal. The parameter $\textrm{SDD}$ is plotted in figure \ref{fig:StatisticalConvergence}(\textit{a}) indicating that throughout the vertical range probed by the LiDAR, the statistical stationarity of the velocity signals can be assumed considering a threshold for the $\textrm{SDD}$ parameter of $30\%$ \citep{Foken2004}.

The convergence of the mean velocity is then analyzed for increasing number of samples \citep{Heisel2018}. The results of this analysis are reported in figure \ref{fig:StatisticalConvergence}(\textit{b}), which shows a good level of statistical convergence as the entire sampling period is used. Finally, the convergence of higher-order statistical moments is qualitatively investigated by inspecting the probability density function of the velocity signal, $P_u(U)$, for the highest range gate, pre-multiplied by $u^p$, where $p$ is the order of the considered central statistical moment. If the tails of the considered function smoothly taper towards zero, then the respective statistical moment, $K_p$, can be considered as adequately estimated through the available data \citep{Meneveau2013}. For the present study, this analysis is performed considering velocity bins with a width of $0.1~\text{m}~\text{s}^{-1}$. In figure \ref{fig:StatisticalConvergence}(\emph{c}), the results suggest a good convergence for the second-order statistics and an incomplete convergence for higher-order statistical moments. 

\subsection{Streamwise mean velocity and turbulence intensity}

\begin{figure}
    \centerline{\includegraphics[width=\textwidth]{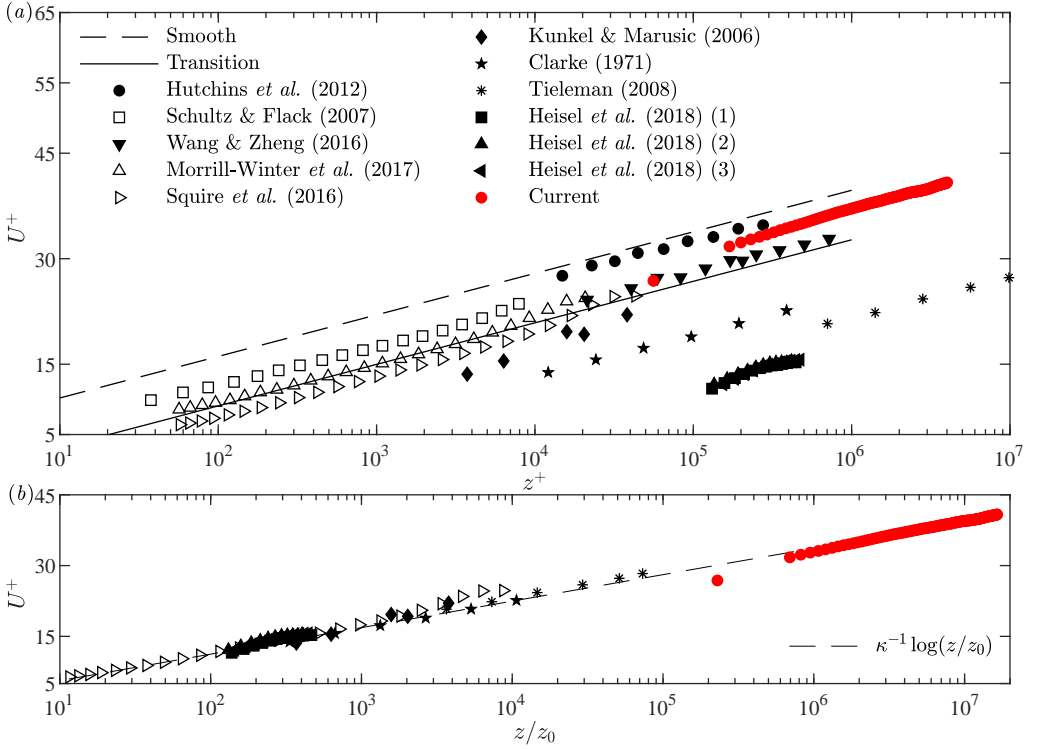}}
    \caption{Mean streamwise velocity measured from the LiDAR and the PA2 sonic anemometer (the lowest point): (\textit{a}) Mean velocity versus inner-scaled wall-normal coordinate; (\textit{b}) Wall-normal coordinate is made non-dimensional with the aerodynamic roughness length, $z_0$. Empty and filled symbols refer to wind tunnel and ASL studies, respectively. This figure is adapted from \cite{Heisel2018}.}\label{fig: MeanVelocity}
\end{figure}

The mean streamwise velocity measured through the wind LiDAR and the PA2 sonic anemometer is scaled with the friction velocity retrieved from the sonic anemometer data, then compared in figure \ref{fig: MeanVelocity}(\emph{a}) to ASL data collected from previous experiments at different sites with variable terrain roughness \citep{Kunkel2006,Tieleman2008,Hutchins2012,Wang2016,Heisel2018}, and laboratory experiments as well \citep{Schultz2007,Squire2016,Morrill2017}. For an ASL flow, the effects of the terrain roughness on the mean streamwise velocity can be accounted through the aerodynamic roughness length, $z_0$, into the logarithmic law of the wall \citep{Kunkel2006,Gryning2007,Metzger2007,Heisel2018}:
\begin{equation}\label{eq: MeanVelocity}
    U^+ = \dfrac{1}{\kappa}\left[\log \left(\dfrac{z}{z_0}\right) - \Psi\left(\dfrac{z}{L} \right) \right],
\end{equation}
where $\Psi$ is the stability correction function \citep{Businger1971}. The experimental data are fitted with (\ref{eq: MeanVelocity}) to estimate the aerodynamic roughness length, which results in $z_0=8.71\times10^{-6}$ m. By normalizing the wall-normal position with $z_0$, a very good agreement is observed in figure \ref{fig: MeanVelocity}(\textit{b}) between the stability-corrected mean streamwise velocity measured by the LiDAR and previous data sets. Furthermore, (\ref{eq: MeanVelocity}) is used to assess the value of $U_\tau$ calibrated on the LiDAR data ($U_\tau=0.51\pm0.009$ m/s) against that estimated from the sonic anemometer data using the eddy-covariance method ($U_\tau=0.42$ m/s).

Another way to account for the terrain roughness on the mean streamwise velocity profile is through the sand-grain roughness parameter, $k_s^+$:
\begin{equation}\label{eq: MeanVelocity_ks}
    U^+ = \dfrac{1}{\kappa}\log\left(\dfrac{z^+}{k_s^+} \right)+B(k_s^+),
\end{equation}
where $B(k_s^+)$ is a function accounting for the vertical shift of the mean velocity profile. For a transitional roughness regime, where $2.25\leq k_s^+\leq90$ \citep{Ligrani1986,Hutchins2012}, $B(k_s^+)$ is given by \citep{Kunkel2006}:
\begin{equation}
    B(k_s^+) = \dfrac{1}{\kappa}\log k_s^+ + 5.0 + \sin\left(\dfrac{\pi h}{2} \right)\left(8.5-5.0-\dfrac{1}{\kappa}\log k_s^+ \right),
\end{equation}
where:
\begin{equation}
    h = \dfrac{\log(k_s^+/90)}{\log(k_s^+/2.25)}.
\end{equation}
Comparing (\ref{eq: MeanVelocity}) and (\ref{eq: MeanVelocity_ks}), the sand-grain roughness can be estimated from $z_0$ as $k_s^+ = 11.4$ ($k_s=0.41$ mm), which is of the same order of magnitude as for previous estimates for the SLTEST site, e.g.  $k_s^+\approx34$ ($k_s=2.9$ mm) in \cite{Kunkel2006}, or  $k_s^+=15$ in \cite{Huang2021}.

\begin{figure}
        \centerline{\includegraphics[width=\textwidth]{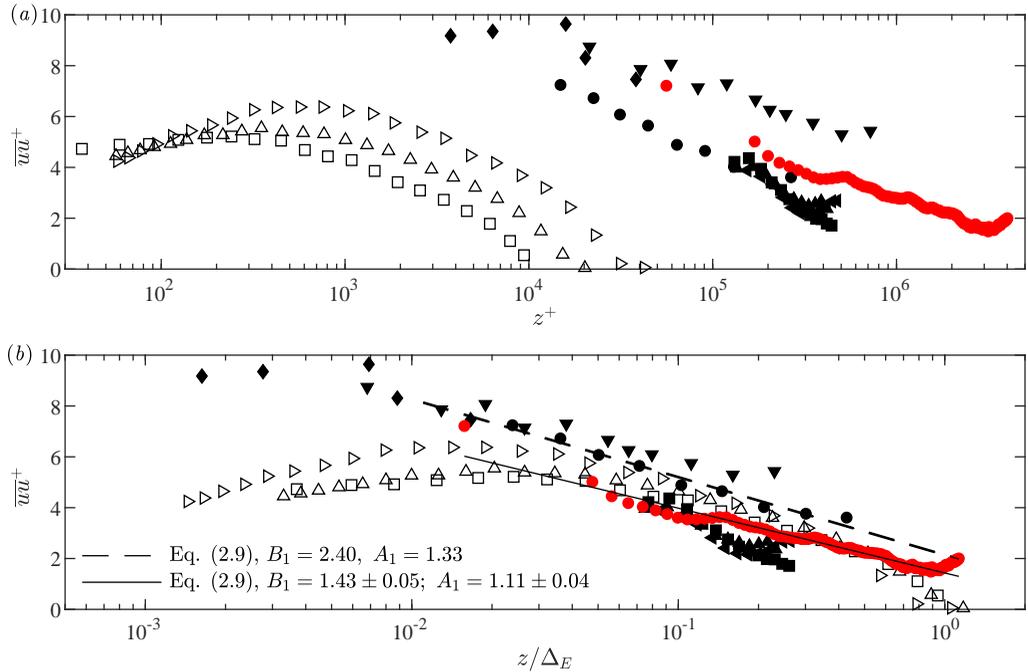}}
    \caption{Wall-normal profile of turbulence intensity with inner- or outer-scaled wall-normal coordinate (plot (\textit{a}) and (\textit{b}), respectively). For the current data set, the lowest point is retrieved from the PA2 sonic anemometer. In panel (\textit{b}) the black continuous line refers to the model of \cite{Marusic2013} calibrated on the present data set. Legend as for figure \ref{fig: MeanVelocity}. This figure is adapted from \cite{Heisel2018}.}\label{fig: Variance}
\end{figure}

The inner-scaled wall-normal profile of the turbulence intensity is reported against the viscous- and outer-scaled wall-normal coordinate in figure \ref{fig: Variance}(\textit{a}) and (\textit{b}), respectively. The estimate of the outer scale of turbulence, $\Delta_E=127$ m for the present data set, is detailed in Appendix \ref{subsec: SpectralGap}. Based on the AEH, the law for the wall-normal distribution of streamwise turbulence intensity can be written as \citep{Townsend1976,Perry1982}: 
\begin{equation}\label{eq: DiagReynStresses}
    {\overline{uu}^+}=B_1-A_1\log\left(\dfrac{z}{\Delta_E}\right), 
\end{equation}
where $B_1$ is a flow-dependent constant accounting for the large-scale inactive motion, while $A_1$ is the Townsend–Perry constant \citep{Perry1986,Baars2020b}. For the SLTEST site under neutral conditions, \cite{Marusic2013} reported: $A_1=1.33\pm0.17$ and $B_1=2.14\pm0.40$, while for the current data set we obtain $A_1=1.11\pm0.04$ and $B_1=1.43\pm0.05$. 

\section{Contribution of eddies with different typology to the streamwise velocity energy}\label{sec: Theory}
\subsection{Reynolds stresses and isolated-eddy function}

According to the AEH, a wall-attached eddy has a wall-normal extent, $\delta $, growing linearly with the distance from the wall, $z$ \citep{Townsend1976,Perry1982}. Therefore, the probability density function representing the occurrence of an eddy with size $\delta$, $p_H(\delta)$, should decrease monotonically with $z$ \citep{Townsend1976}: 
\begin{equation}\label{eq:pdf}
    p_H(\delta) = \left\{\begin{array}{cc}
         \dfrac{M}{\delta}&~\textrm{as}~\delta_1\leq\delta\leq\Delta_E  \\
         0&~\textrm{otherwise} 
    \end{array},\right.
\end{equation}
where $M$ is a constant related to the eddy population density on the plane of the wall \citep{Desilva2015a}, and $\delta_1\approx100\nu/U_{\tau}$ is the smallest eddy size owning to the logarithmic layer, which is fixed by the viscous cutoff \citep{Kline1967,Perry1982}. The Reynolds stresses at a given $z$ are then calculated as weighted integral of isolated-eddy contributions over the entire scale range:
\begin{equation}\label{eq:TotalStress}
    \overline{u_iu_j}^+=\int_{\delta_1}^{\Delta_E}I_{ij}\left( \dfrac{z}{\delta}\right)p_h(\delta)~\text{d}\delta=\int_{\delta_1}^{\Delta_E}MI_{ij}\left( \dfrac{z}{\delta}\right)\dfrac{\text{d}\delta}{\delta} = \int_{z/\Delta_E}^{z/\delta_1}MI_{ij}\left(\dfrac{z}{\delta} \right)\dfrac{\delta}{z} \text{d}\left(\dfrac{z}{\delta}\right),
\end{equation}
where the function $I_{ij}$ is referred to as “eddy function”, representing the geometrically self-similar isolated-eddy contribution to $\overline{u_iu_j}^+$. In the view of the AEH, $I_{ij}$ is determined by the sole geometrical features of the archetypal wall-attached eddy. Remarkably, $I_{ij}$ can also be estimated for a fixed eddy size, $\delta$, through the differential form of (\ref{eq:TotalStress}):
\begin{equation}\label{eq: I_ij}
		I_{ij}\left(\dfrac{z}{\delta} \right)=-\dfrac{z}{M} \dfrac{\partial \overline{u_i u_j}^+}{\partial z}.
\end{equation}
The term on the right-hand side of (\ref{eq: I_ij}) is commonly referred to as “indicator function” and it has been used to detect the presence and extent of the logarithmic region \citep[e.g.][]{Bernardini2014,Lee2015,Hwang2018,Yamamoto2018,Klewicki2021}. It is noteworthy that equation (\ref{eq: I_ij}) represents the contribution to the Reynolds stresses of the sole wall-attached eddies (type $\mathcal{A}$), and, thus, does not encompass the contribution of wall-detached eddies, or coherent structures characterized by larger wavelengths, e.g. VLSMs and superstructures (type $\mathcal{B}$). 

\subsection{Regions of the streamwise velocity energy spectrum}\label{subsec: Hogstrom}
Regarding the streamwise velocity spectrum, as mentioned in \S\ref{sec: Intro}, for wall-normal locations owning to the inertial layer, the non-dimensional low-wavenumber limit of the $k_x^{-1}$ region \citep[denoted as $F$ following the notation of][]{Perry1986} scales with $\Delta_E$ \citep{Perry1982,Perry1986}. Therefore, the large eddies with scale $O(\Delta_E)$ will contribute to the streamwise velocity energy spectrum as:

\begin{equation} \label{eq: PHC86_Delta}
    \phi_{uu}^+(k_x\Delta_E) = g_1(k_x\Delta_E) = \dfrac{\phi_{uu}(k_x)}{\Delta_EU_\tau^2}.
\end{equation}
On the other hand, the non-dimensional high-wavenumber limit of the inverse-power-law spectral region, $P$, scales with $z$, and the respective eddies contribute to the streamwise velocity energy spectrum as:

\begin{equation} \label{eq: PHC86_z}
    \phi_{uu}^+(k_xz) = g_2(k_xz) = \dfrac{\phi_{uu}(k_x)}{zU_\tau^2}.
\end{equation}
Considering an overlapping region where equations (\ref{eq: PHC86_Delta}) and (\ref{eq: PHC86_z}) hold simultaneously, and, thus, equating $\phi_{uu}(k_x)$ from (\ref{eq: PHC86_Delta}) and (\ref{eq: PHC86_z}), we obtain:

\begin{equation} \label{eq: kx_PHC86}
     \dfrac{g_1(k_x\Delta_E)}{g_2(k_xz)}= \dfrac{z}{\Delta_E}.
\end{equation}
Therefore, within this overlapping region, $g_1$ and $g_2$ must be of the form \citep{Perry1975,Perry1986,Davidson2009}:
 
 \refstepcounter{equation}\label{eq: A1_PHC86}
$$
  g_1(k_x\Delta_E) = \dfrac{A_1}{k_x\Delta_E}; \quad
  g_2(k_xz) = \dfrac{A_1}{k_xz},
  \eqno{(\theequation{\mathit{a},\mathit{b}})}\label{eq: Inverse_PHC86}
$$
where $A_1$ is the Townsend-Perry constant, which is of the order of $1$ \citep{Perry1986,Baars2020b}.
The turbulence intensity associated with wall-attached eddies is then expressed as integral of the streamwise velocity energy spectrum over the different regions:
\begin{equation}\label{eq: uu_PHC86}
    \overline{uu}^+ = \int_0^{F} g_1(k_x\Delta_E)~\text{d}(k_x\Delta_E) + \int_{Fz/\Delta_E}^P g_2(k_xz)~\text{d}(k_xz) + \int_{P}^\infty \phi_{uu}^+(k_xz)\text{d}(k_xz).
\end{equation}

\begin{figure}
        \centerline{\includegraphics{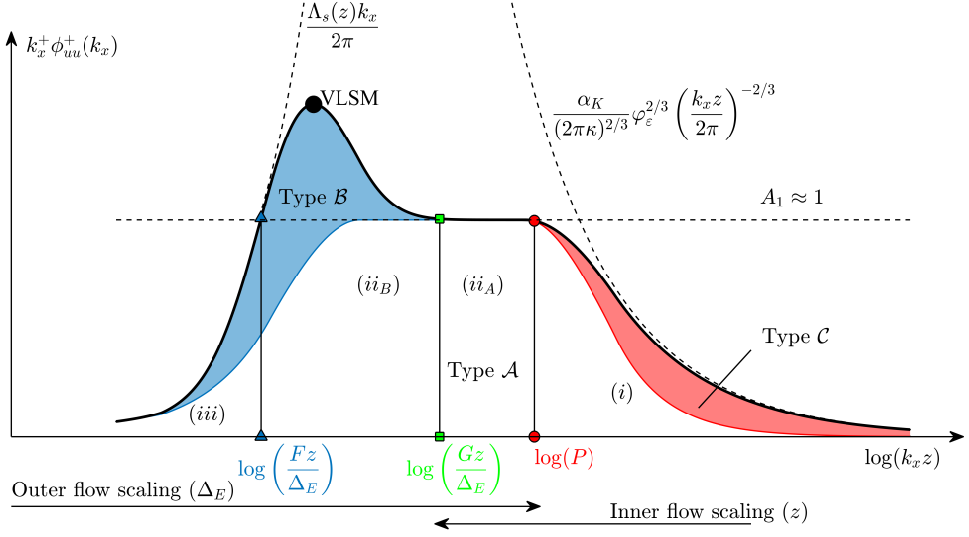}}
    \caption{Sketch of the different regions encompassed in the streamwise velocity pre-multiplied energy spectrum.}\label{fig: SpectrumExample}
\end{figure}

A similar approach for the identification of different regions of the streamwise velocity energy spectrum was proposed by \cite{Hogstrom2002} for ASL flows.  Specifically, three different regions are singled out within the turbulence spectral range by this model, which are indicated with dashed lines in the sketch reported in figure \ref{fig: SpectrumExample}. Region ($i$) is the inertial subrange, which follows the Kolmogorov law:
\begin{equation}\label{eq: Hog_i}
    k_x^+\phi_{uu}^+ (k_x)= \dfrac{\alpha_K}{(2\pi\kappa)^{2/3}}\varphi_{{\varepsilon}}^{2/3} \left(\dfrac{k_xz}{2\pi} \right)^{-2/3},
\end{equation}
where $\alpha_K$ is the Kolmogorov constant, $\kappa$ is the von K\'{a}rm\'{a}n constant and $\varphi_{{\varepsilon}}$ is the non-dimensional dissipation rate, which can be estimated as follows \citep{Kaimal1972}: 

\begin{equation}\label{eq: phi_epsilon}
    \varphi_\varepsilon^{2/3}\left(\dfrac{z}{L}\right)=\left(\dfrac{\kappa z \varepsilon}{U_\tau^3}\right)^{2/3}=\left\{ 
         \begin{array}{ll}
         1+0.5\left|\dfrac{z}{L} \right|^{2/3},&~-2\leq z/L\leq0\\
         1+2.5\left|\dfrac{z}{L} \right|^{3/5},&~0\leq z/L\leq2\
         \end{array}
         \right.,
\end{equation} 
where $\varepsilon$ is the turbulent kinetic energy dissipation rate and $L$ is the Obukhov length (see \S\ref{subsec: LiDARfield}). It is noticed that the maximum value attained by the stability correction function in (\ref{eq: phi_epsilon}) is equal to 1.32 at $z=143$ m for the data set under investigation.

Region ($ii$) corresponds to the spectral range where the pre-multiplied spectrum is nearly constant:
\begin{equation}\label{eq: Hog_ii}
    k_x^+\phi_{uu}^+ (k_x)\approx A_1,
\end{equation}
where $A_1$ is the Townsend-Perry constant in (\ref{eq: DiagReynStresses}). The wall-normal range where region ($ii$) was observed in the ASL, which is dubbed as “eddy surface layer” (ESL), has a thickness of about $\Delta_E/3$ \citep{Hunt2000,Hogstrom2002}. This estimate is similar to that for the height of the logarithmic layer for ASL flows reported in \cite{Hutchins2012}, while \cite{Marusic2013} conservatively quantified the height of the logarithmic layer based on mean velocity profiles and turbulence intensity at $z=0.15\Delta_E$ for laboratory flows.

For region ($iii$), the pre-multiplied spectrum increases with the wavenumber:

\begin{equation}\label{eq: Hog_iii}
    k_x^+\phi_{uu}^+ (k_x)= \Lambda_s \dfrac{k_x}{2\pi}.
\end{equation}
The parameter $\Lambda_s$ is a large-scale characteristic wavelength estimated as $\Lambda_s=A(z)U_\tau/f_C$, where $f_C$ is the Coriolis frequency \citep{Rossby1935} ($f_C=9.38\times10^{-5}$Hz for the present data set), the parameter $A$ is linearly proportional to $z$ within the ESL, while $\Lambda_s$ reaches the maximum value at the height of the ESL, then decreases \citep{Hogstrom2002}. 

The model of \cite{Hogstrom2002} defined through (\ref{eq: Hog_i},~\ref{eq: Hog_ii}) and (\ref{eq: Hog_iii}) allows for calculating the boundaries of region ($ii$). Specifically, from $A(z)$ and $A_1$, the non-dimensional low-wavenumber limit of region (${ii}$), $F$, can be found equating (\ref{eq: Hog_ii}) and (\ref{eq: Hog_iii}):

\begin{equation}\label{eq: F_kx_l}
    F=\dfrac{2\pi f_CA_1\Delta_E}{A(z)U_\tau}.
\end{equation}
Similarly, the wavenumber at the intersection between region ($i$) and (${ii}$) is found equating (\ref{eq: Hog_i}) and (\ref{eq: Hog_ii}):

\begin{equation}\label{eq: P_kx_u}
    P=\dfrac{1}{\kappa}\left(\dfrac{\alpha_K\varphi_\varepsilon^{2/3}}{A_1} \right)^{3/2}.
\end{equation}

Building upon the spectral model proposed by \cite{Hogstrom2002}, we propose to further divide region (${ii}$) into two high- and low-wavenumber sub-regions referred to as (${ii}_A$) and (${ii}_B$), respectively, where the energy contribution of wall-attached eddies is predominant for the former, while that associated with VLSMs is predominant for the latter (see figure \ref{fig: SpectrumExample}). This approach is inspired by previous works; for instance, \cite{Kim1999} identified two distinct peaks in the streamwise energy spectrum for pipe flows associated with VLSMs and LSMs. Similarly, \cite{Rosenberg2013} and \cite{Vallikivi2015} modeled separately the VLSM spectral peak through a Gaussian function (here associated with region ($ii_B$)) and the flat pre-multiplied region through a cubic spline. For ASL flows, \cite{Wang2016} calculated the energy fraction associated with either VLSMs or LSMs partitioning the streamwise energy spectrum into wavelength intervals, i.e. $\lambda_x>3\Delta_E$ for VLSMs and $0.3\Delta_E<\lambda_x<3\Delta_E$ for LSMs. For our study, the non-dimensional wavenumber at the intersection between regions ($ii_A$) and ($ii_B$), indicated with $G$ in figure \ref{fig: SpectrumExample}, is thought to scale with $\Delta_E$ considering that its spectral value is affected by the energy content of VLSMs, which scale with $\Delta_E$. Finally, the maximum height where region ($ii_A$) can be observed, $z_\textrm{max}$, is identified where the condition $P=z_\textrm{max}G/ \Delta_E$ is fulfilled.

The technical strategy for the quantification of $F,~G$, and $P$ from the pre-multiplied streamwise velocity spectrum is detailed in the following. Starting from region ($i$), the term $\alpha_K/\kappa^{2/3}$ in (\ref{eq: Hog_i}) is fitted by overlapping  the pre-multiplied streamwise velocity spectra versus the inertia-scaled wavenumber for all the heights probed by the LiDAR and over the frequency range $k_xz\geq 2$. The fitting of the experimental spectra leads to an estimate of  $\alpha_K/\kappa^{2/3}=0.60$ for the present data set. 

For region ($ii_A$), $A_1$ is heuristically determined for each height in the proximity of the spectral range where the high-frequency limit $P$ is expected ($k_xz\approx1$). Subsequently, the intersection between the horizontal line equal to $A_1$ and the fitted spectrum for region ($i$) leads to the identification of the high-wavenumber limit $P$.

For region ($iii$), $A(z)$ is obtained for each height through the fitting of the streamwise velocity energy spectra with (\ref{eq: Hog_iii}) limited to the low-frequency energy-increasing portion. Then, for each height, the intersection between the horizontal line equal to $A_1$ and the fitted spectrum for region ($iii$) identifies the low-frequency limit $F$. Finally, the inner boundary between regions ($ii_A$) and ($ii_B$), $G$, is heuristically quantified at the crossing between the energy-decreasing region for smaller wavenumbers typically associated with VLSMs and the horizontal line equal to $A_1$. 

\subsection{Identification of the energy associated with different eddy typology based on the linear coherence spectrum of the streamwise velocity}\label{subsec: Baars}
The scale-dependent cross-correlation of two statistically stationary velocity signals collected at wall-normal positions $z$ and $z_R$ (reference height) can be estimated 
through the two-point linear coherence spectrum (LCS):
\begin{equation}\label{eq: LCS}
    \gamma^2(z,z_R;k_x) = \dfrac{|\phi_{uu}'(z,z_R;k_x)|^2}{\phi_{uu}(z;k_x)~\phi_{uu}(z_R,k_x)},
\end{equation}
where $||$ indicates the modulus while $\phi_{uu}'(z,z_R;k_x)$ is the cross-spectral density of the two streamwise velocity signals, which is practically the Fourier transform of the cross-variance function between $u(z)$ and $u(z_R)$. Therefore, the LCS represents the fraction of common variance shared by $u(z_R)$ and $u(z)$ across frequencies \citep{Bendat1986}. Due to the normalization with the single-point energy spectra, we have $0\leq\gamma^2\leq1$. Considering that the LCS is calculated from the amplitude of the cross-spectral density, no information is retained about the phase shift of the shared energy between the two velocity signals \citep{Nelson2013,Baars2017}. 

\begin{figure}
    \centerline{\includegraphics{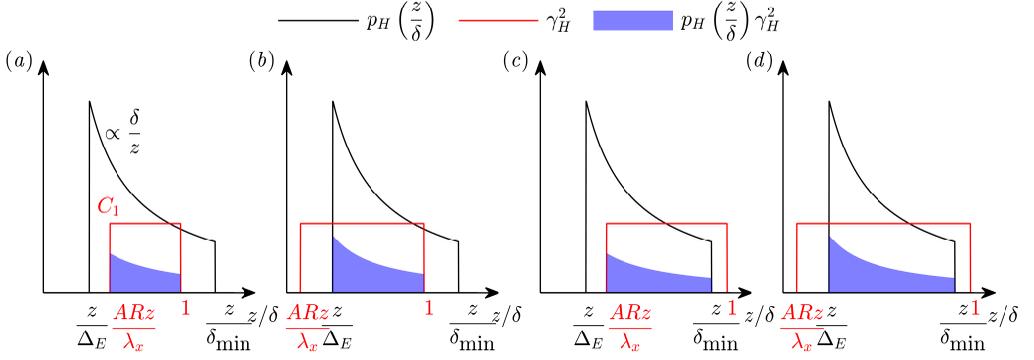}}
    \caption{Isolated-eddy contribution to the LCS, $\gamma_H^2$ (\ref{eq: gamma_H_l}), p.d.f. of an isolated eddy with vertical extent $\delta$, $p_H$, and their product for the various combinations of $\Delta_E,~\delta_\textrm{min}$ and $\AR$.}\label{fig: LCS_cases}
\end{figure}

Considering a boundary layer flow encompassing only wall-attached eddies generated by a single hierarchy with wavelength $\lambda_H$, vertical size $\delta$, and, thus, aspect ratio $\AR = \lambda_H / \delta$, the non-zero portion of the LCS is limited to wall-normal positions with $z\leq\delta$, since no wall-attached eddies are present above, and to wavelengths with $\lambda_x\geq\lambda_H=\AR \delta$ due to the concatenation and random repetitions of the same hierarchy along the streamwise direction \citep{Baars2017}. Therefore, the isolated-eddy contribution to the LCS for a single hierarchy can be modeled as:  
\begin{equation}\label{eq: gamma_H_l}
    \gamma_{H}^2 \left(\dfrac{\lambda_x}{\delta},\dfrac{z}{\delta} \right) = C_0~ H \left[1-\dfrac{z}{\delta} \right]H \left[\dfrac{\lambda_x}{\AR\delta}-1 \right];~\text{as}~\dfrac{z}{\delta}\leq 1,~\dfrac{\lambda_x}{\AR\delta}\geq 1,
\end{equation}
where $H$ is the unit Heaviside function, while the parameter $C_0$ ($0<C_0<1$) represents the isolated-eddy contribution to the LCS, and it only depends on the geometric features of the archetypal eddy. When a continuous distribution of attached eddies is considered, the resulting LCS will then be expressed as the sum of the various isolated contributions weighted by their probability density function throughout the scale range:
\begin{equation}\label{eq: gamma_l}
    \gamma^2\left(\dfrac{\lambda_x}{\Delta_E};\dfrac{z}{\delta_\textrm{min}},\dfrac{z}{\Delta_E} \right) = \min\left[\int_{\delta_\textrm{min}}^{\Delta_E}M~\gamma_{H}^2 \left(\dfrac{\lambda_x}{\delta},\dfrac{z}{\delta} \right) \dfrac{\text{d}\delta}{\delta},1\right],
\end{equation}
where $\delta_\textrm{min}$ is equal to $\delta_1$ or $z_R$ if the latter owns to the near-wall or logarithmic region, respectively. Considering that:
\begin{equation*}
    H\left[\dfrac{\lambda_x}{\AR\delta
    }-1 \right] = H\left[\dfrac{z}{\delta}-\dfrac{\AR z}{\lambda_x} \right],
\end{equation*}
equation (\ref{eq: gamma_l}) becomes:
\begin{equation}\label{eq: gamma_l_zdelta}
    \gamma^2\left(\dfrac{\lambda_x}{\Delta_E};\dfrac{z}{\delta_\textrm{min}},\dfrac{z}{\Delta_E} \right) = \min\left\{\int_{z/\Delta_E}^{z/\delta_\textrm{min}}C_1 H \left[1-\dfrac{z}{\delta} \right]H\left[\dfrac{z}{\delta}-\dfrac{\AR z}{\lambda_x} \right]\dfrac{\delta}{z}\text{d}\left(\dfrac{z}{\delta} \right),1\right\},
\end{equation}
where $C_1=MC_0$. Therefore, for a certain $z$ and $\lambda_x$, the region where the contribution to the LCS is non-zero depends on the values of $\Delta_E,~\delta_\textrm{min}$ and $\AR$. Specifically, four combinations are possible among these limits, each of them schematically reported in figure \ref{fig: LCS_cases}. For instance, assuming a wall-normal location within the logarithmic region ($z\geq \delta_1$) and $\AR z\leq\lambda_x\leq\AR \Delta_E$, the case reported in figure \ref{fig: LCS_cases}(\emph{a}) is obtained, where the active boundaries of the non-zero contribution to the LCS are $\AR z/\lambda_x$ and $1$. Thus, equation (\ref{eq: gamma_l_zdelta}) becomes:
\begin{equation}
    \gamma^2=\min\left\{\int_{\AR z/\lambda_x}^{1}C_1 \dfrac{\delta}{z} \text{d}\left( \dfrac{z}{\delta}\right),~1\right\} = \min\left\{C_1\log\left(\dfrac{\lambda_x}{\AR z} \right),~1 \right\},
\end{equation}
which is the model proposed by \cite{Baars2017}. This case with the three remaining combinations between active boundaries of (\ref{eq: gamma_l_zdelta}) (sketched in figure \ref{fig: LCS_cases}\emph{b}-\emph{d}) lead to the analytical formulation of the LCS for an attached-eddy flow.

In \cite{Baars2017}, it was noted that the LCS becomes zero for heights below $\Delta_E$. To encompass this feature, an offset for $\gamma^2$, denoted as $C_3$, is added to the analytical formulation of the LCS model developed from (\ref{eq: gamma_l_zdelta}), which then reads as:
\begin{equation}\label{eq: LCS_cases}
    \gamma^2 = \left\{ \begin{array}{ll}
      \min\left[C_1\log \left(\dfrac{\lambda_x}{\AR z} \right),1\right]~&\text{if}~ z\geq z_R,~\lambda_x/\Delta_E \leq \exp\left(\dfrac{C_3}{C_1}\right)\AR~(a) \\[8pt]
       \min\left[-C_1\log \left(\dfrac{z}{\Delta_E} \right)+C_3,1\right]~&\text{if}~ z\geq z_R,~\lambda_x/\Delta_E> \exp\left(\dfrac{C_3}{C_1}\right)\AR~(b) \\[8pt]
        \min\left[C_1\log \left(\dfrac{\lambda_x}{\AR\delta_\textrm{min}} \right),1\right]~&\text{if}~ z< z_R,~\lambda_x/\Delta_E\leq \exp\left(\dfrac{C_3}{C_1}\right)\AR~(c) \\[8pt]
        \min\left[C_1\log \left(\dfrac{\Delta_E}{\delta_\textrm{min}} \right)+C_3,1\right]~&\text{if}~ z< z_R,~\lambda_x/\Delta_E> \exp\left(\dfrac{C_3}{C_1}\right)\AR ~(d)\
    \end{array}\right.    .
\end{equation}
The threshold wavelength $\lambda_x^{th}/\Delta_E = \exp(C_3/C_1)\AR$ is obtained enforcing continuity in $\lambda_x$ between (\ref{eq: LCS_cases}\textit{a}) and (\ref{eq: LCS_cases}\textit{b}), and it represents the boundary between wall-attached eddies and coherent structures generated from their streamwise concatenation. It is noteworthy that including $C_3$ implies $\gamma^2=0$ at $z_{max}/\Delta_E=\exp(C_3/C_1)$, rather than at $z/\Delta_E=1$. In \cite{Baars2017}, $\gamma^2=0$ was identified for $z\gtrapprox0.7\Delta_E$ while the authors estimated $C_1=0.302$, which leads to $C_3=-0.103$.  

All the cases in (\ref{eq: LCS_cases}) were observed experimentally in \cite{Baars2020a}, and it is shown here that they are consistent with a continuous distribution of wall-attached eddies. In particular, assuming $z_R$ within the near-wall region, we have $\delta_\textrm{min}=\delta_1$ and (\ref{eq: LCS_cases}) becomes equation (4.7) of \cite{Baars2020a}. Similarly, assuming $z_R$ within the logarithmic region we have $\delta_\textrm{min}=z_R$, which leads to equation (4.10) of \cite{Baars2020a}. 

It is noteworthy that in \cite{Baars2020a} $C_1$ and $\AR$ are calibrated twice from the experimental distribution of LCS depending on whether the reference height was located in the near-wall or the logarithmic region. Specifically, it was estimated $C_1=0.302$ and $\AR=14.01$ if $z_R$ resides in the near-wall region and $C_1=0.383$ and $\AR=13.18$ if $z_R$ resides in the  logarithmic region. This discrepancy is attributed by \cite{Baars2020a} to the use of a single convection velocity for the application of the \cite{Taylor1938} hypothesis for frozen turbulence. From the proposed LCS model, we confirm that $C_1$ should not change with the reference height assuming a constant $\AR$.

\subsection{Notes on the linear coherence spectrum of the streamwise velocity}\label{subsec: LCS_notes}
By considering the logarithmic coordinates $\lambda^*=\log(\lambda_x/\Delta_E)$ and $z^*=\log(z/\Delta_E)$, the LCS model of (\ref{eq: LCS_cases}) can be rewritten, by only considering the case with $\gamma^2\leq1$, as follows:
\begin{equation}\label{eq: LCS_cases STAR}
    \gamma^2 = \left\{ \begin{array}{ll}
      C_1\left(\lambda^*-\lambda^*_{th}\right) -C_1 \left(\Delta z^* + z^*_R \right) +C_3~&\text{if}~ \Delta z^*\geq 0,~\lambda^* \leq \lambda^*_{th}~(a) \\[8pt]
      -C_1 \left(\Delta z^* + z^*_R \right) +C_3~&\text{if}~ \Delta z^*\geq 0,~\lambda^* > \lambda^*_{th}~(b) \\[8pt]
      C_1\left(\lambda^*-\lambda^*_{th}\right) -C_1 z^*_R +C_3~&\text{if}~ \Delta z^*< 0,~\lambda^* \leq \lambda^*_{th}~(c) \\[8pt]
        -C_1 z^*_R +C_3~&\text{if}~ \Delta z^*< 0,~\lambda^* > \lambda^*_{th}~(d)\\[8pt]
    \end{array}\right.    ,
\end{equation} 
where $\lambda^*_{th}=C_3/C_1+\log(\AR)$ and $\Delta z^*=\log(z/z_R)$. Based on (\ref{eq: LCS_cases STAR}), the following considerations can be made:
\begin{enumerate}
    \item for $\lambda^* \leq \lambda^*_{th}$, $\gamma^2$ is a linear function of $\lambda^*$, its slope is not a function of $z^*$ and has a constant value equal to $C_1$;
    \item for $\lambda^* > \lambda^*_{th}$, $\gamma^2$ achieves an asymptotic value, $\gamma_\infty^2$, which is a linear function of $\Delta z^*$ with slope $-C_1$ if $\Delta z^*\geq0$ while it is constant for $\Delta z^*<0$;
    \item the threshold wavelength, $\lambda^*_{th}$, is not a function of $z^*$;
    \item the intercept of $\gamma^2$ for $\lambda^* \leq \lambda^*_{th}$ is equal to the asymptotic value achieved for $\lambda^* > \lambda^*_{th}$ for every $z^*$.
\end{enumerate}

By imposing $\gamma^2=0$ in (\ref{eq: LCS_cases}), we obtain the boundary conditions for a non-null LCS:
\begin{equation}\label{eq: LCS_zero}
    \left\{ \begin{array}{ll}
      z^*=\lambda^*-\log(\AR)~&\text{if}~ \Delta z^*\geq 0,~\lambda^* \leq \lambda^*_{th}~(a) \\[8pt]
      z^*=C_3/C_1 ~&\text{if}~ \Delta z^*\geq 0,~\lambda^* > \lambda^*_{th}~(b) \\[8pt]
      \lambda^*=z_R^*+\log(\AR)~&\text{if}~ \Delta z^*< 0,~\lambda^* \leq \lambda^*_{th}~(c) \\[8pt]
      z_R^*=C_3/C_1~&\text{if}~ \Delta z^*< 0,~\lambda^* > \lambda^*_{th}~(d)\\[8pt]
    \end{array}\right.    ,
\end{equation} 
The relationship (\ref{eq: LCS_zero}\emph{a}) represents the line separating the coherent from the non-coherent component of the flow with $z_R$ in the ($\lambda_x$, $z$) domain. According to the LCS model, this line represents the spectral boundary between region ($i$) and region ($ii_A$) for wall-normal positions above $z_R$. Below $z_R$, the zero-contour of $\gamma^2$ has a constant value of $\lambda_x$ (\ref{eq: LCS_zero}\emph{c}) because, according to the AEH, eddies with a height smaller than $z_R$ do not contribute to the LCS. It is noteworthy that the spectral boundary between region ($i$) and region ($ii_A$) is only a function of $\AR$, and it shifts towards larger wavelengths as $\AR$ increases.

The remaining relationships obtained by imposing $\gamma^2=0$ provide $z_{max}$ for $\lambda^* > \lambda^*_{th}$ in (\ref{eq: LCS_zero}\emph{b}), and the trivial condition that $z_R$ should be lower than $z_{max}$ to achieve a non-null $\gamma^2$  (\ref{eq: LCS_zero}\emph{d}).

Similarly, we can estimate the conditions for the saturation on the contribution of wall-attached eddies to the LCS by imposing  $\gamma^2=1$ in (\ref{eq: LCS_cases}):
\begin{equation}\label{eq: LCS_saturation}
    \left\{ \begin{array}{ll}
      z^*=\lambda^*-\log(\AR)-1/C_1~&\text{if}~ \Delta z^*\geq 0,~\lambda^* \leq \lambda^*_{th}~(a) \\[8pt]
      z^*=(C_3-1)/C_1 ~&\text{if}~ \Delta z^*\geq 0,~\lambda^* > \lambda^*_{th}~(b) \\[8pt]
      \lambda^*=z_R^*+\log(\AR)+1/C_1~&\text{if}~ \Delta z^*< 0,~\lambda^* \leq \lambda^*_{th}~(c) \\[8pt]
      z_R^*=(C_3-1)/C_1~&\text{if}~ \Delta z^*< 0,~\lambda^* > \lambda^*_{th}~(d)\\[8pt]
    \end{array}\right.    ,
\end{equation} 
From (\ref{eq: LCS_saturation}\emph{a}), it is noted that the line demarcating the LCS saturation above $z_R$ is parallel to that separating the spectral region ($i$) from ($ii_A$), yet translated towards larger wavelengths by $\Delta \lambda^*=1/C_1$. The same shift applies below $z_R$ (\ref{eq: LCS_saturation}\emph{c}), which provides an insightful physical interpretation for $C_1$, namely $C_1$ controls the spectral width where the energy contributions due to wall-attached eddies build up. Furthermore, equations (\ref{eq: LCS_saturation}\emph{b}) and (\ref{eq: LCS_saturation}\emph{d}) provide the maximum wall-normal position and the maximum $z_R$ where LCS saturation is achieved for $\lambda^* > \lambda^*_{th}$. Further, the range in the wall-normal position where the LCS is non-null is equal to $\Delta z^*=1/C_1$. 

\section{Detection of the energy associated with eddies of different typology from LiDAR measurements}\label{sec: Results}
\subsection{Energy spectra of the streamwise velocity}\label{subsec: Spectra}
As detailed in Appendix \ref{subsec: SpectralGap}, the spectral gap between mesoscales and turbulent scales is identified at a frequency $f_\text{gap}=0.0055$ Hz, while the outer scale of turbulence is estimated as $\Delta_E=127$ m. Therefore, the friction Reynolds number for the data set under investigation is equal to $\Rey_{\tau}=U_{\tau}\Delta_E/\nu=3.55\times10^6$. The streamwise velocity signals collected through the wind LiDAR and the sonic anemometer are high-pass filtered with a cutoff frequency equal to $f_\text{gap}$ through the filter proposed by \cite{Hu2020} to isolate only velocity fluctuations connected with turbulence motions. 

\begin{figure}
    \centerline{\includegraphics{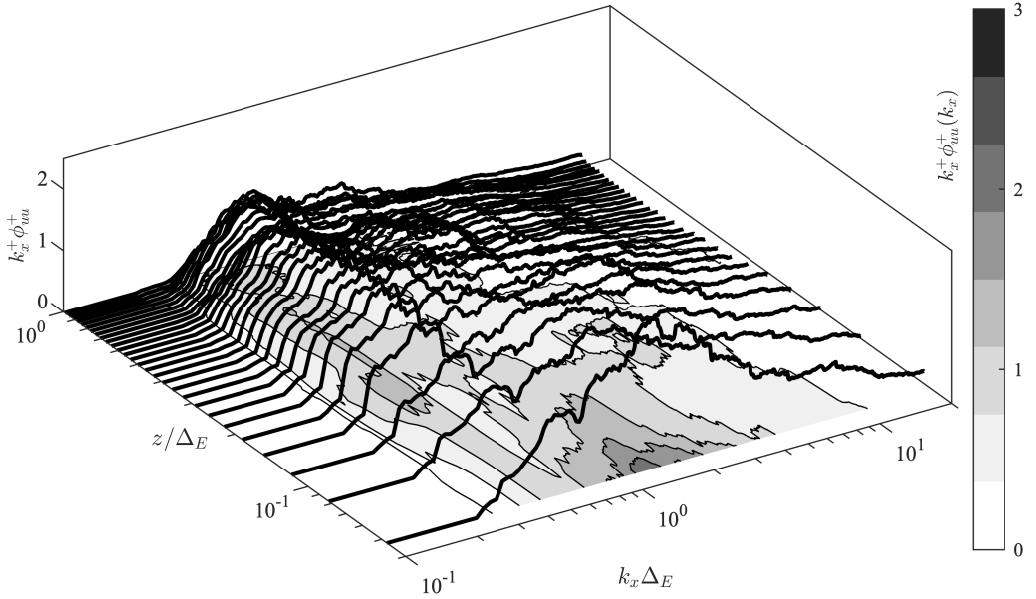}}
    \caption{Pre-multiplied energy spectra of the streamwise velocity obtained from the wind LiDAR measurements.}\label{fig: Spectra3D}
\end{figure}

The streamwise velocity energy spectra are calculated through the Welch algorithm \citep{Welch1967} using a window size equal to $0.0003$ Hz and $10\%$ overlapping. Each spectrum is evaluated over $N/2+1$ frequencies linearly spaced between 0 and the Nyquist frequency ($0.5$ Hz for the LiDAR signals and $10$ Hz for the sonic anemometer measurements), where $N$ is the total number of samples of the velocity signals ($3580$ and $72000$ for LiDAR and sonic-anemometer data, respectively).
The spectra are then smoothed using a moving average applied at each frequency $f_i$ using a variable spectral stencil of $f_i\pm 0.35f_i$ \citep{Baars2020a}. The obtained streamwise-velocity pre-multiplied energy spectra are reported in figure \ref{fig: Spectra3D} versus the outer-scaled wavenumber and the wall-normal distance. The wavenumber is calculated as $k_x=2\pi f/U(z)$ by leveraging the Taylor's hypothesis of frozen turbulence \citep{Taylor1938}, where $f$ is frequency. 

\begin{figure}
    \centerline{\includegraphics{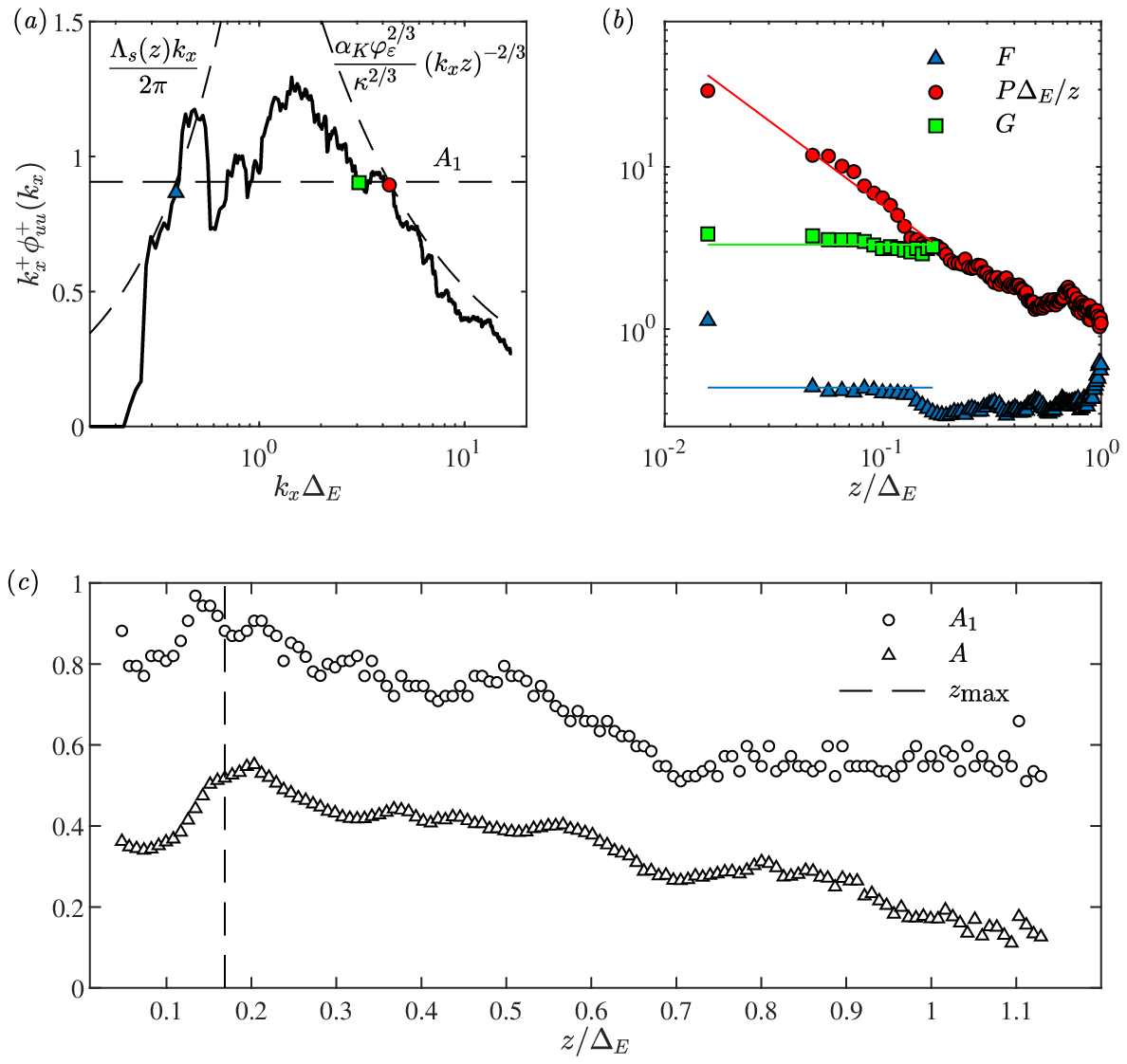}}
    \caption{Identification of spectral limits from the streamwise velocity energy spectra: (\emph{a}) Detection of the spectral limits for the LiDAR velocity signal collected at a 16-m height; (\emph{b}) Wall-normal profiles of $F$, $G$ and $P$; (\emph{c}) Vertical profiles of $A_1$ and $A$.}\label{fig: BestFit_A}
\end{figure}

As mentioned in \S\ref{subsec: Hogstrom}, from the streamwise velocity pre-multiplied spectra calculated at each height, $A_1$ is estimated heuristically, while the parameter $A(z)$ is obtained by fitting the spectrum with (\ref{eq: F_kx_l}) over the spectral region ($iii$). An example of this procedure is reported in figure \ref{fig: BestFit_A}(\emph{a}) for the pre-multiplied streamwise velocity spectrum evaluated for the LiDAR data collected at a height of 16 m. A spectral region with roughly constant energy, i.e. corresponding to region ($ii_A$), is identified in the proximity of the low-frequency limit of region ($i$) ($k_x\Delta_E\approx5$). The fitting of (\ref{eq: Hog_i}) for the model of region ($i$) with the experimental spectrum is reported with a dashed line over the high-frequency part of the spectrum, while the fitting with (\ref{eq: Hog_iii}) for the model of region ($iii$) is also reported with a dashed line over the low-frequency part of the spectrum. The intersections between the horizontal dashed line corresponding to the identified value of $A_1$ ($\approx0.91$ for the LiDAR data reported in figure \ref{fig: BestFit_A}\emph{a}) and the modeled spectra for regions ($i$) and ($iii$) enable the identification of the non-dimensional spectral limit $P$ (red circle in figure \ref{fig: BestFit_A}\emph{a}) and $F$ (blue triangle in figure \ref{fig: BestFit_A}\emph{a}), respectively. Finally, for the identification of the spectral limit $G$, the part of region ($ii$) with energy larger than $A_1$ yet reducing with increasing wavenumber is considered. The spectral limit $G$ is associated with the intersection of this part of the spectrum with the horizontal dashed line corresponding to $A_1$.

The vertical profiles of the spectral limits $P$, $G$ and $F$ obtained from the analysis of the pre-multiplied streamwise velocity energy spectra are reported in figure \ref{fig: BestFit_A}(\emph{b}). First, it is observed that, as predicted from the AEH (see \S\ref{sec: Intro}), the non-dimensional spectral limits $G$ and $F$ are roughly invariant with $z$. Specifically, $G$ has a mean value of about 3.3, which corresponds to a wavelength of $1.9\Delta_E$, while $F$ has a mean value of 0.44, which corresponds to a wavelength of $14.3\Delta_E$. In contrast, the spectral limit between region ($i$) and region ($ii_A$), $P$, decreases with the wall-normal position, which confirms its inertial-scaling consistently with the predictions of \cite{Perry1982,Perry1986} and the experimental results of \cite{Nickels2005,Hwang2015,Baars2017,Hu2020} for laboratory flows. The linear fitting of the profile reported in figure \ref{fig: BestFit_A}(\emph{b}), produces an estimate for $P$ of 0.58, which corresponds to an eddy aspect ratio of $\lambda_x/z=2\pi/0.58\approx10.8$, which is smaller than the value of 14 estimated in \cite{Baars2017} through the LCS analysis of data sets collected with a wide range of Reynolds number flows, including an ASL case as well.

As mentioned in \S\ref{subsec: Hogstrom}, the intersection between $P\Delta_E/z$ and $G$ identifies the maximum height where region (${ii}_A$) is observed, i.e. $z_\mathrm{max}\approx21$ m ($0.17\Delta_E$). It is noteworthy that this outer-scaled value coincides with that provided by \cite{Hwang2015} and it is close to the results by \cite{Baars2020b} ($0.15\Delta_E$), while it is lower than the value proposed by \cite{Hu2020} ($0.53\Delta_E$). 

Besides the limits amongst the different spectral regions, it is important to analyze the vertical profile of the parameter $A_1$, which is heuristically estimated from the pre-multiplied spectra of the streamwise velocity. In the perspective of the AEH, $A_1$ can be estimated only for wall-normal positions lower than $z_\mathrm{max}$, namely for heights where the spectral region ($ii_A$) can be detected. However, in this work, a value is still associated with $A_1$ even aloft, which corresponds to the average energy within the spectral range between regions ($i$) and ($iii$). The obtained wall-normal profile of $A_1$ is reported in figure \ref{fig: BestFit_A}(\emph{c}). Starting from the lowest height and moving upwards, $A_1$ increases from a value of about 0.8 up to about 1 at $z/\Delta_E\approx0.13$. While the obtained values are reasonable according to previous works (e.g., $A_1=0.80$ for channel flows and $A_1=1.0$ for ASL and boundary layers \citep{Hu2020}, while $A_1=0.975$ for boundary layers \citep{Baars2020b}), the variability of $A_1$ with height is not in agreement with the AEH predictions. Further, for wall-normal positions above $z_\mathrm{max}$, the mean energy roughly monotonically reduces up to $z/\Delta_E\approx0.7$, then it remains roughly constant aloft. 

An interpretation of these experimental results might be provided by analyzing the vertical profile of the parameter $A(z)$ of (\ref{eq: F_kx_l}), which represents the energy level over region ($iii$) as a function of height and it is connected with the vertical variability of the turbulent kinetic energy associated with large-scale turbulent motions. Figure \ref{fig: BestFit_A}(\emph{c}) shows that $A(z)$ increases from 0.37 up to 0.60 between $0.1\Delta_E$ and $0.2\Delta_E$, followed by a roughly monotonic decrease aloft. These values are comparable to those reported by \cite{Hogstrom2002}, namely $A(z)$ increases from 0.2 up to 1 at a height of $z=0.3\Delta_E$, then it decreases aloft. 
The similar trends obtained for $A_1$ and $A(z)$ as a function of the wall-normal position suggest that even though a roughly flat region of the pre-multiplied spectra can be singled out over the region ($ii_A$), the values attained by $A_1$ can significantly be affected by an underlying energy contribution associated with VLSMs, which is a reasonable feature within region ($ii$) where wall-attached and coherent structures with larger wavelengths co-exist and, thus, their energy contributions overlap.  
\begin{figure}
   \centerline{\includegraphics{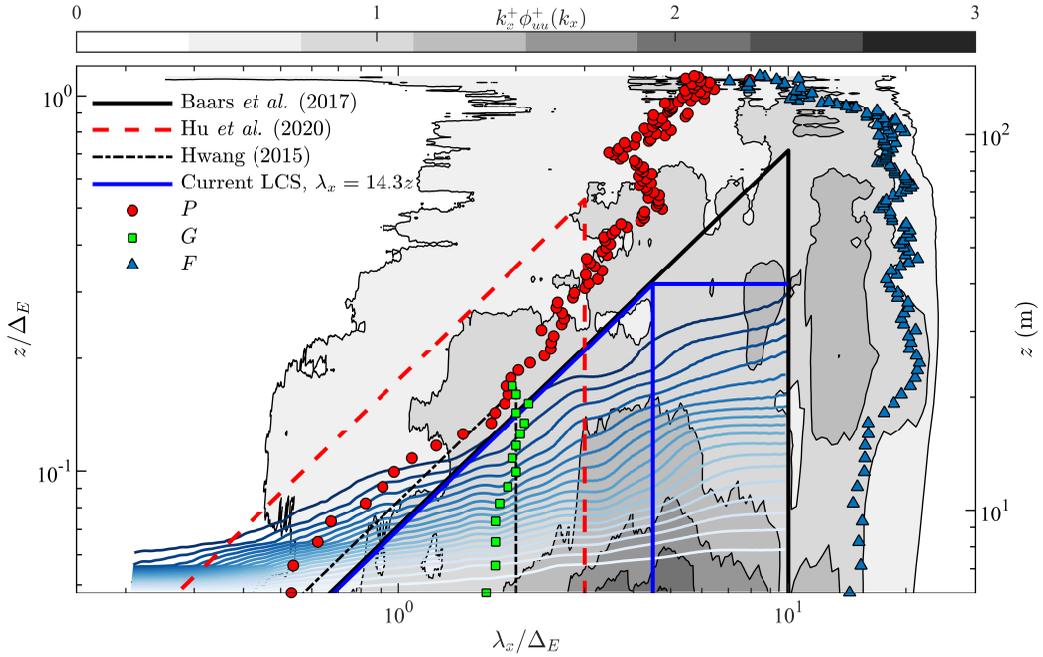}}
    \caption{Pre-multiplied energy spectra of the LiDAR streamwise-velocity measurements reported as a gray colormap (iso-contour levels from 0.4 up to 3 with a 0.4 step). The linear coherence spectra are reported as blue-scaled contour lines (iso-contour levels from 0.1 up to 1 with a 0.05 step).  \label{fig:Spectra2D}}
\end{figure}

To better visualize the boundaries of the various spectral regions, the streamwise velocity pre-multiplied energy spectra are reported in figure \ref{fig:Spectra2D} with a gray colormap. The estimated spectral limits for the different heights, $P$, $G$, and $F$, are reported as well with red circle, green square and blue triangle markers, respectively. The spectral limits of region ($ii$) estimated from the above-mentioned previous works are also reported in that figure. For heights below 0.17$\Delta_E$, the wall-normal trend of the spectral limit $P$ has a good level of agreement with the findings for channel flows by \cite{Hwang2015}, where the aspect ratio of the wall-attached eddies was estimated as $\AR=12$. The present results for $P$ are slightly larger than the spectral limit estimated through the LCS in \cite{Baars2017}, where the authors estimated $\AR=14$ for boundary layers. 

From the analysis of the spectral limit $G$ between regions ($ii_A$) and ($ii_B$) (green squares in figures \ref{fig: BestFit_A}(\emph{b}) and \ref{fig:Spectra2D}) the average value of $\lambda_x=1.9\Delta_E$ is close to that obtained by \cite{Hwang2015} ($\lambda_x=2\Delta_E$). Finally, for the spectral limit between region ($ii$) and ($iii$), the outer-scaled wavelength associated to $F$ (blue triangles in figures \ref{fig: BestFit_A}(\emph{b}) and \ref{fig:Spectra2D}) is quantified as $14.3\Delta_E$ below $z_\textrm{max}$ and $15.8\Delta_E$ above, both within the interval provided by \cite{Hutchins2012} for the maximum streamwise extent of VLSMs in a boundary layer ($10\div20\Delta_E$). Finally, the comparison for the spectral limits of region ($ii$) obtained from the current investigation with those from previous studies \citep{Hwang2015,Baars2017,Hu2020} are summarized in table \ref{tab: FP}.

\begin{table}
\begin{center}
    \begin{tabular}{lcccccc}
Reference             &  Flow type & $z_\textrm{max}/\Delta_E$                                                                   & Limits $k_x^{-1}$ region & $P$      & $G$   & $F$    \\ [3pt]
\cite{Baars2017}      &  TBL, CH, ASL   & $-$            &$14z\leq \lambda_x\leq 10\Delta_E$    & 0.45 & $0.63$ & -  \\[1pt]
\cite{Hu2020}         & TBL, CH, ASL    & $0.53$           &$5.7z\leq \lambda_x\leq 3\Delta_E$    & 1.10 & 2.09 & $-$  \\[1pt]
\cite{Hwang2015}      &  CH             & $0.17$  &$12z\leq \lambda_x\leq 2\Delta_E$     & 0.52 & 3.14 & $-$ \\[1pt]
\cite{Nickels2005}    & TBL             & $0.02$  &$15.7z\leq \lambda_x\leq 0.3\Delta_E$ & 0.40 & 20.94 & $-$  \\[1pt]
Present spectra        & ASL           & $0.17$  & $10.8z\leq\lambda_x\leq1.9\Delta_E$   & 0.58   & 3.30 & 0.44 \\
Present LCS        & ASL           & $0.31$  & $14.3z\leq\lambda_x\leq4.5\Delta_E$   & 0.44   & 1.40 & $-$ \\
\end{tabular}
\caption{Identification of the various regions from the energy spectra and linear coherence spectrum of the streamwise velocity. The acronyms TBL, CH, and ASL stand for turbulent boundary layer, channel flow, and atmospheric surface layer, respectively. \label{tab: FP}}
\end{center}
\end{table}

\subsection{Detection of different energy contributions in the LiDAR measurements through the linear coherence spectrum}\label{subsec: LCS}

The energy contributions associated with different eddy typologies are further investigated through the analysis of the LCS, which is calculated according to (\ref{eq: LCS}) through the \cite{Welch1967} algorithm and leveraging as reference the velocity signal collected from the lowest LiDAR gate  ($z_{R}\approx0.05\Delta_E$). Specifically, each signal is subdivided into windows with a 179-s duration (corresponding to nearly $18 \Delta_E$ leveraging the Taylor's hypothesis of frozen turbulence \citep{Taylor1938}) and $90\%$ overlapping between consecutive windows to estimate the auto- and cross-spectra of the streamwise velocity. It is noticed that the values for the window duration and overlapping of the various data subsets do not distort the LCS at lower frequencies, while these parameters are more important for ensuring a roughly null LCS at higher frequencies. Each LCS spectrum is smoothed as performed for the energy spectra. For more details on the calculation of the LCS and determination of its parameters see Appendix \ref{app: Smoothing}. The LCS calculated from the LiDAR measurements is reported as blue iso-contours over the ($\lambda_x , z$)-domain in figure \ref{fig:Spectra2D}.

\begin{figure}
    \centerline{\includegraphics[width=\textwidth]{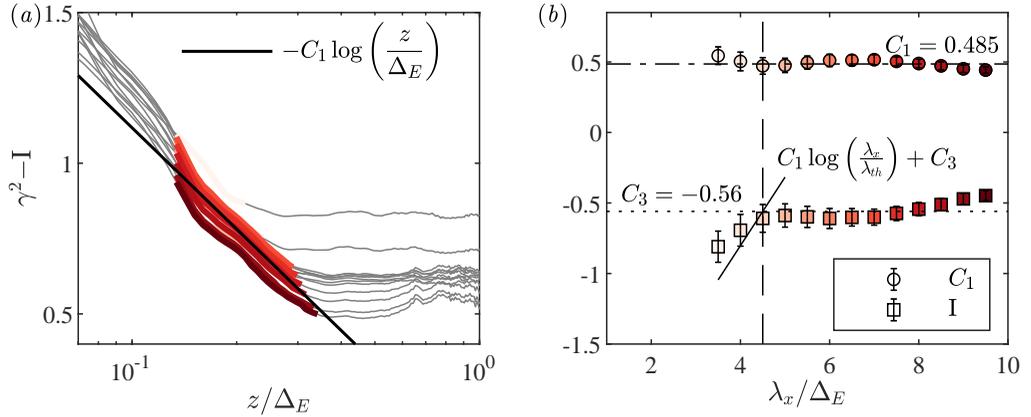}}
    \caption{Linear coherence spectrum of the LiDAR measurements using $z_{R}/\Delta_E\approx0.05$: (\emph{a}) $\gamma^2-I$ for fixed values of $\lambda_x$; (\emph{b}) Fitted values of $C_1$ and $I$ from (\ref{eq: LCS_cases STAR}); error bars refer to $95\%$ confidence level. The parameter $\lambda_x^{th}$ is reported with a vertical dashed line. \label{fig: LCS_infty}}
\end{figure}

The parameters of the analytical LCS model for wall-attached eddies (\ref{eq: LCS_cases}) are quantified by fitting the LCS calculated from the LiDAR measurements. According to the point ($ii$) discussed in \S\ref{subsec: LCS_notes}, for a fixed $\lambda_x$, $\gamma^2$ is a linear function of $\log(z/z_R)$ with a slope equal to $-C_1$ for heights above $z_R$. To this aim, $\lambda_x$ is selected within the interval between $3.5\Delta_E$ and $9.5\Delta_E$ to avoid effects due to small-wavelength eddies and large-wavelength non-turbulent contributions, respectively. The portions of $\gamma^2$ varying roughly linearly with $\log(z/z_R)$, which are reported with a red color scale in figure \ref{fig: LCS_infty}(\emph{a}), are fitted with a linear function of $\log(z/z_R)$ to estimate the intercept $I$ and the slope $-C_1$, which are then reported in figure \ref{fig: LCS_infty}(\emph{b}) with square and circle markers, respectively. $C_1$ has roughly constant values for different $\lambda_x$, as predicted from the analytical model for the LCS of (\ref{eq: LCS_cases STAR}), with an average value of $C_1=0.485$. 

For $z\geq z_R$, the analytical LCS model predicts that the intercept of $\gamma^2$, $I$, should be a linear function of $\log(\lambda_x)$ for $\lambda_x\leq\lambda_x^{th}$, while achieving an asymptotic value equal to $C_3$ for $\lambda_x>\lambda_x^{th}$ (\ref{eq: LCS_cases STAR}). In figure \ref{fig: LCS_infty}(\emph{b}), the values of $I$ confirm these predictions, specifically showing an asymptotic value of the intercept of $C_3\approx-0.56$ for $\lambda_x/\Delta_E\geq4.5$, thereby assumed as $\lambda_x^{th}$. An assessment of the results obtained with this fitting procedure is performed considering that the intercept of $\gamma^2$ should vary as $C_1\log(\lambda_x/\lambda_x^{th})+C_3$ for $\lambda_x\leq\lambda_x^{th}$ and $z\geq z_R$. This function, which is reported in figure \ref{fig: LCS_infty}(\emph{b}) with a black solid line using the fitted values obtained for $C_1$ and $C_3$, and $\lambda_x^{th}$ shows a good agreement with the experimental data considering the small number of samples available for $\lambda_x\leq\lambda_x^{th}$.

Recalling that $\log(\lambda_x^{th}/\Delta_E)=C_3/C_1+\log(\AR)$, we can estimate the aspect ratio of the coherent eddies as $\AR\approx14.3$. Finally, the maximum height reached by wall-attached eddies is estimated from (\ref{eq: LCS_zero}\emph{b}), namely $z_{max}/\Delta_E=\exp(C_3/C_1)\approx0.31$. This estimate of $z_{max}$ obtained from the LCS analysis is larger than that obtained from the analysis of the energy spectra, i.e. $0.17\Delta_E$ (\S\ref{subsec: Spectra}), as well as the estimates obtained by \cite{Hwang2015} and \cite{Baars2020b} ($0.17\Delta_E$ and $0.15\Delta_E$, respectively); furthermore, the spectral approach of \cite{Hu2020} returns a sensibly larger value of $z_{max}$ ($0.53\Delta_E$). Finally, in \cite{Baars2017}, for an aspect ratio of 14, the authors estimated $\lambda_x^{th}=10\Delta_E$ based on $z_{max}=0.71\Delta_E$. 

The estimate of $z_{max}/\Delta_E\approx0.31$ obtained from the LCS analysis agrees with the vertical extent of the eddy surface layer (ESL) for ASL flows, which was previously estimated equal to $0.3\Delta_E$ \citep{Hunt2000,Hogstrom2002,Drobinski2007}. In the existing literature, the ESL is considered as the layer where the vertical confinement induced by the wall affects the dynamics and evolution of eddies entraining the boundary layer from aloft that, in turn, contribute to the generation of new turbulence structures from the wall. Our analysis seems to indicate that the ESL is dominated by a hierarchical distribution of eddies statistically attached to the wall.

The calibrated LCS model, and specifically the $\gamma^2=0$-condition (\ref{eq: LCS_zero}\emph{a}), is now compared with the experimental values of $\gamma^2$, as reported in the color map of figure \ref{fig:Spectra2D}. Analyzing the darkest iso-contour with $\gamma^2=0.1$, a deviation from the predicted unitary slope of the LCS iso-contours is observed for $z\lessapprox0.12\Delta_E$, which is due to the local contribution of small, wall-incoherent structures (type $\mathcal{C}$-eddies) \citep{Krug2019,Baars2020a} as a consequence of the small vertical separation between $z_R$ and the wall-normal position of these LiDAR data. Moving towards higher vertical positions and for $\lambda_x/\Delta_E\approx2$, a unitary slope of the isocontours is recovered, in agreement with the LCS model (\ref{eq: LCS_zero}\emph{a}), and in agreement with previous experimental results \citep{Baars2017,Baars2020a,Li2021}. For higher wall-normal positions, a further deviation from the model prediction is observed, which may be ascribed to a residual thermal stratification still present in the ASL flow considering the early-morning time of the data collection (local time between 3:00 AM and 4:00 AM), as already observed from previous field experiments \citep{Krug2019}.

It is noteworthy that the zero-LCS contour predicted with (\ref{eq: LCS_zero}\emph{a}) has a relatively good agreement with the vertical profile of the spectral limit $P$ between region ($i$) and ($ii_A$) identified through the analysis of the streamwise velocity energy spectra in \S\ref{subsec: Spectra} (figure \ref{fig:Spectra2D}), yet translated towards slightly larger wavelengths. An aspect ratio of 14.3 is estimated indeed from the LCS analysis, while $\AR\approx10.8$ from the analysis of the energy spectra. Nonetheless, this result can be considered as proof that the inverse-power-law region of the streamwise velocity energy spectra is associated with wall-attached eddies \citep{Perry1986}, and the LCS approach resonates with the spectral approach to identify the $k_x^{-1}$ high-frequency limit.

The LCS analytical model calibrated through the LiDAR data for the high-frequency limit (\ref{eq: LCS_zero}\emph{a}) is in good agreement with the analysis based on the LCS presented in \cite{Baars2017} for boundary layers ($\AR=14$), while a slightly smaller aspect ratio ($\AR=12$) was estimated in \cite{Hwang2015} for channel flows. 
In contrast, the high-frequency spectral limit provided in \cite{Hu2020} with $\AR=5.7$ seems to be an underestimate for the present data set.

Regarding the spectral boundary between the energy associated with wall-attached eddies and that due to their streamwise concatenation, e.g. VLSMs and superstructures, the spectral limit $G$ estimated through the analysis of the streamwise velocity energy spectra is about 1.9$\Delta_E$, which is very similar to the respective limit estimated by \cite{Hwang2015}, and is significantly lower than the value of 4.5$\Delta_E$ obtained from the present LCS analysis through $\lambda_x^{th}$. For the sake of completeness, $G$ was estimated equal to 2.09$\Delta_E$ in \cite{Hu2020} and 10$\Delta_E$ in \cite{Baars2017} (estimated indirectly from $\AR$ and $z_{max}$). This analysis would suggest that the analysis of streamwise velocity energy spectra should lead to an underestimate of the spectral limit $G$ due to the co-existence of energy contributions associated with eddies of different typologies. In contrast, the present LCS approach should offer a more reliable approach to separate wall-attached spectral energy from that associated with larger coherent structures.

Finally, the present LCS method does not provide a criterion to identify the low-frequency limit of the spectral region ($ii_B$), $F$, which is then only identifiable through the analysis of the streamwise velocity energy spectra.

\begin{figure}
    \centering
    \includegraphics[width=\textwidth]{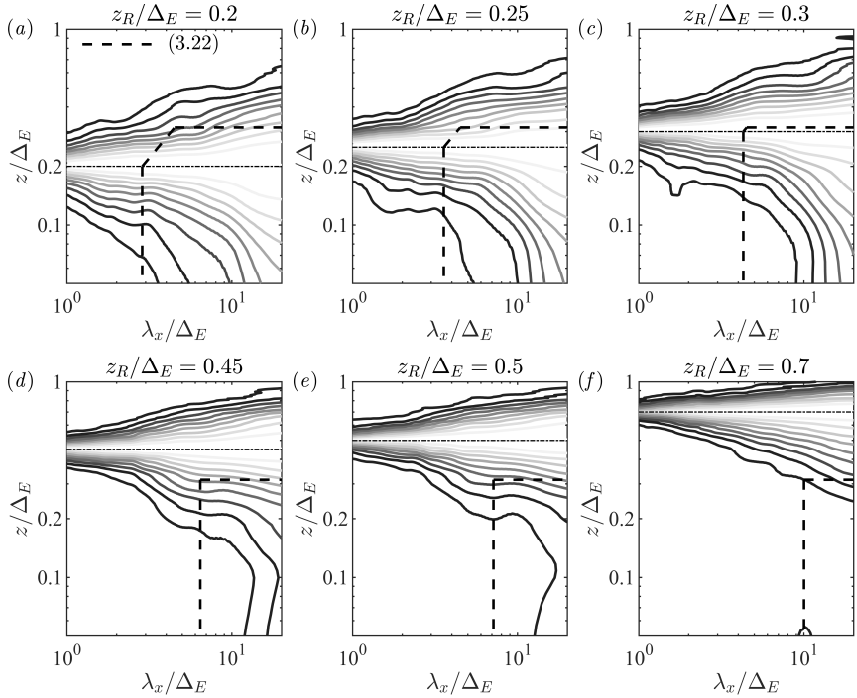}
    \caption{Iso-contours (values from 0.1 up to 1 with a step of 0.1) of the linear coherence spectrum obtained with different $z_R$ (horizontal dot-dashed lines). The $\gamma^2=0$-contours predicted through the model in (\ref{eq: LCS_zero}) are reported with a dashed line.  
    \label{fig: LCS_height}}
\end{figure}

\subsection{Linear coherence spectrum calculated with increasing reference height}\label{subsec: LCS_height}
After the LCS analysis performed with the reference height $z_R=0.05\Delta_E$, a similar analysis is carried out by increasing $z_R$, which can provide more insights into the LCS analytical model (\ref{eq: LCS_cases}), especially for $z<z_R$. However, we do not expect this analysis to further contribute to the foregoing discussion for $z>z_R$ because, with increasing $z_R$, the vertical range where incremental contributions associated with wall-attached eddies can be observed, $z_{maz}-z_R$, reduces. To this aim, six further reference heights are selected, namely $0.20\Delta_E,~0.25\Delta_E,~0.30\Delta_E,~0.45\Delta_E,~0.50\Delta_E$ and $0.70\Delta_E$. 

The LCS maps obtained for all the considered $z_R$ values are reported as isocontours in figure \ref{fig: LCS_height}; further, the $\gamma^2=0$-isocontours estimated analytically through (\ref{eq: LCS_zero}) are also reported using the parameters $C_1$, $C_3$, and $\AR$ calibrated as for \S\ref{subsec: LCS}. For the region at smaller wavelengths demarcated by the analytically-predicted $\gamma^2=0$ isocontours, it is evident that the experimental LCS obtained from the LiDAR measurements do not follow the analytical model of (\ref{eq: LCS_cases}), rather they are dominated by energy contributions associated with wall-detached type-$\mathcal{C}$ eddies \citep{Krug2019,Baars2020a}. However, with increasing $z_R$, thus increasing the vertical distance between $z_R$ and the lower LiDAR sampling heights, which is advantageous to reduce the effects of type-$\mathcal{C}$ eddies on the LCS, a roughly vertical $\gamma^2=0$ isocontour is observed already for $z_R/\Delta_E=0.2$, which is even more evident for $z_R/\Delta_E=0.25$ and $0.3$. As predicted through the LCS analytical model (\ref{eq: LCS_zero}\emph{c}), for $z<z_R$, the $\gamma^2=0$-isocontour should occur for $\lambda_x=\AR~z_R$, which is indeed in good agreement for all the $z_R$ values used for this analysis below $z_{max}$ by assuming $\AR=14.3$ as estimated in \S\ref{subsec: LCS}. The vertical $\gamma^2=0$-isocontour becomes less evident increasing $z_R$, until it disappears completely for $z_R/\Delta_E=0.7\Delta_E$, indicating that no structures originating from below $z_R=0.7$ are coherent with that reference position.

\begin{figure}
    \centering
    \includegraphics[width=\textwidth]{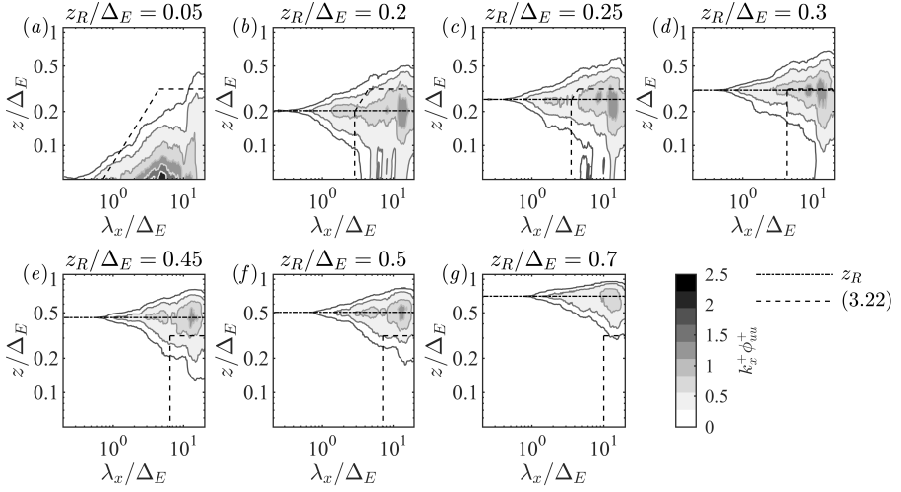}
    \caption{Coherent portion of the streamwise velocity energy spectra calculated for different reference heights, $z_R$. \label{fig: LCS_height_spectra}}
\end{figure}

\begin{figure}
    \centering
    \includegraphics[width=\textwidth]{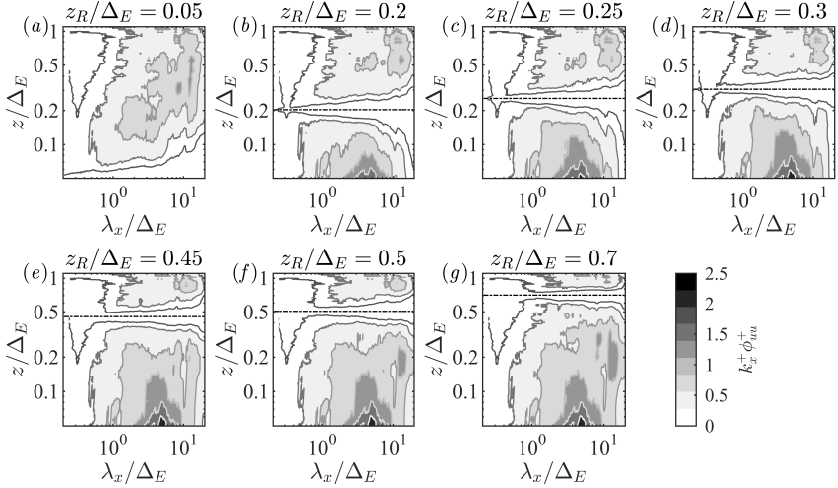}
    \caption{Incoherent portion of the streamwise velocity energy spectra calculated for different reference heights, $z_R$. \label{fig: LCS_height_spectra_inc}}
\end{figure}

The maps of the streamwise velocity energy spectra for the coherent (figure \ref{fig: LCS_height_spectra}) and incoherent (figure \ref{fig: LCS_height_spectra_inc}) components calculated for all the selected reference heights are now analyzed. Starting from the lowest reference height, $z_R=0.05\Delta_E$, it is observed that the coherent energy is practically confined within the $\gamma^2=0$ limit predicted in (\ref{eq: LCS_zero}\emph{a}) and the maximum height (\ref{eq: LCS_zero}\emph{b}) (figure \ref{fig: LCS_height_spectra}\emph{a}). The energy peak is observed at the lowest height ($z_R=0.05\Delta_E$) at a wavelength $\lambda_x/\Delta_E\approx 5.2$. This is practically the upper region of the outer energy peak already observed by, e.g., \cite{Wang2016} for atmospheric flows.

On the other hand, figure \ref{fig: LCS_height_spectra_inc}(\emph{a}) shows that the incoherent component obtained using $z_R=0.05\Delta_E$ encompasses energy across the entire spectral range considered and up to $\Delta_E$. Furthermore, the energy seems to move towards larger wavelengths with increasing height, which is a similar feature predicted from the AEH for wall-attached eddies. This suggests that also wall-detached eddies might be affected by the wall confinement of the flow and the local shear in the boundary layer. The incoherent energy component may be thought of as the footprint of shear surface layer (SSL) structures entraining from above via top-down motions \citep{Hunt2000,Hogstrom2002,Morrison2007}, thus incoherent with a reference height located in the eddy surface layer (ESL). 

With increasing reference height, i.e. for $z_R=0.2\Delta_E$, the coherent energy below $z_R$ (figure \ref{fig: LCS_height_spectra}\emph{b}), drastically reduces compared to the case with $z_R=0.05\Delta_E$ (figure \ref{fig: LCS_height_spectra}\emph{a}), and the remaining coherent energy is mainly located towards large wavelengths ($O(\lambda_x/\Delta_E)\approx10$). This indicates that a cluster of coherent structures originated below $z_R=0.2\Delta_E$, i.e. wall-attached eddies, does not attain this height, and only taller wall-attached eddies are left in the coherent energy component, which are characterized by larger wavelengths indeed.

For the same reason, increasing the reference height from $0.05\Delta_E$ to $0.2\Delta_E$ leads to enhanced incoherent energy below $z_R$ (compare figure \ref{fig: LCS_height_spectra_inc}\emph{b} with \ref{fig: LCS_height_spectra_inc}\emph{a}). This added energy resembles the coherent energy obtained with $z_R=0.05\Delta_E$ (figure \ref{fig: LCS_height_spectra}\emph{a}). Therefore, this analysis corroborates that the increase of reference height from $0.05\Delta_E$ to $0.2\Delta_E$ mainly leads to transferring a certain energy packet associated with wall-attached eddies from the coherent to the incoherent component.

It is noteworthy that in figure \ref{fig: LCS_height_spectra}(\emph{b}) a significant amount of coherent energy is singled out around $z_R=0.2\Delta_E$, which is the effect on the LCS due to type-$\mathcal{C}$ eddies and their streamwise concatenation. A similar feature is observed for all the reference heights located above $0.05\Delta_E$ (figure \ref{fig: LCS_height_spectra}\emph{b}-\emph{g}).

A similar trend is observed with increasing $z_R$ below $z_{max}\approx 0.31 \Delta_E$, i.e. reduced coherent energy and increased incoherent energy below $z_R$, while above $z_{max}$, i.e. in figure \ref{fig: LCS_height_spectra}(\emph{e}-\emph{g}), the coherent component shows only the contribution associated with type-$\mathcal{C}$ eddies and no energy extending below, indicating no contribution due to wall-attached eddies. On the other hand, above $z_{max}$, the incoherent component achieves practically an asymptotic energy map for $z<z_{max}$, and only energy at large scales for $z>z_{max}$ are added with increasing $z_R$, which confirms that no wall-attached eddies are statistically present above $z_{max}$. 

\subsection{Streamwise turbulence intensity}\label{subsec: BandIntegration}
\begin{figure}
    \centerline{\includegraphics{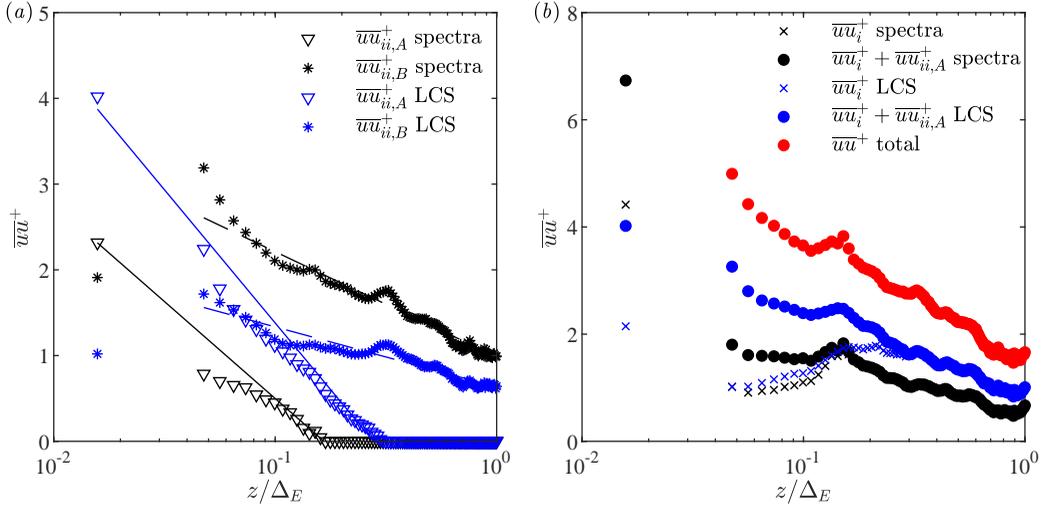}}
    \caption{Vertical profiles of the streamwise turbulence intensity considering different eddy typologies and identification techniques: (\emph{a}) contributions over region ($ii_A$) and ($ii_B$) identified through the spectral method and the linear coherence spectrum (LCS); (\emph{b}) cumulative contributions to $\overline{uu}$.}\label{fig: Variance_AE_DE}
\end{figure}

The energy contributions to the streamwise velocity associated with wall-attached and VLSMs have been identified through the analysis of the streamwise velocity energy spectra (\S\ref{subsec: Spectra}), and the LCS (\S\ref{subsec: LCS}). Considering that the streamwise turbulence intensity is the integrated spectral energy across scales, it is then possible to distinguish the contributions to $\overline{uu}$ associated with eddies of different typologies. The streamwise turbulence intensity obtained using the integration limits estimated either from the spectral analysis (\S\ref{subsec: Spectra}) or the LCS (\S\ref{subsec: LCS}) are reported in figure \ref{fig: Variance_AE_DE}(\emph{a}) with black and blue markers, respectively. Specifically, the streamwise turbulence intensity associated with wall-attached eddies, $\overline{uu}_{ii,A}$, is obtained by integrating the streamwise velocity energy spectra between the limits $P$ and $G$ for the spectral method, while for the LCS method the integration is performed between the limits $\lambda_x=\AR~z$ and $\min[\lambda_x=\AR~z~\exp{1/C_1},~\lambda_x^{th}]$ (\ref{eq: LCS_zero} and \ref{eq: LCS_saturation}). Similarly, $\overline{uu}_{ii,B}$ is obtained by integrating the energy spectra between the limits $G$ and $F$ for the spectral method, and between $\min[\lambda_x=\AR~z~\exp{1/C_1},~\lambda_x^{th}]$ and $\lambda_x/\Delta_E=2\pi/F$, namely using the same low-frequency limit estimated through the spectral method. 

Concerning the model for $\overline{uu}_{ii,A}^+$ derived from the AEH (\ref{eq: DiagReynStresses}), in figure \ref{fig: Variance_AE_DE}(\emph{a}), the data  seemingly show a logarithmic vertical profile for both spectral and LCS methods, even though with significant differences in energy content and maximum height. The wall-attached component $\overline{uu}_{ii,A}$ is lower for the spectral method than for the LCS method, because, as reported in \S\ref{subsec: Spectra},  while the spectral limit $P$ is slightly larger for the spectral method, the spectral limit $G$ is significantly over-estimated with respect to the LCS outcome (shorter spectral range between $P$ and $G$ for the spectral method), due to the overlap of the energy associated with larger coherent structures, e.g. VLSMs and superstructures, concealing the flat part of the pre-multiplied streamwise velocity energy spectra of region ($ii_A$). For the same reason, the component $\overline{uu}_{ii,B}$ is larger for the spectral method than for the LCS method. Similarly, $z_{max}$ estimated from the spectral analysis is lower than that obtained from the LCS analysis being the limit $2\pi/G$ smaller than $\lambda_x^{th}\approx4.5$ estimated through the LCS method, thus the intersection of the high-frequency limit of the spectral region ($ii_A$) with the low-frequency limit occurs at higher wall-normal positions for the results obtained from the LCS.

The vertical profiles of $\overline{uu}^+_{ii,A}$ are fitted with the model based on the AEH in (\ref{eq: DiagReynStresses}) to estimate the parameters $A_1$ and $B_1$, which are reported in table \ref{tab. A1_B1}. It is noteworthy that the fitting has been limited to LiDAR gates with $z\geq18$ m, which is the LiDAR range gate, to avoid possible underestimation of $\overline{uu}$ due to the LiDAR spatial averaging over the probe volume (see Appendix \ref{sec: SpectraCorrection}). As shown in table \ref{tab. A1_B1}, $A_1$ is generally estimated very close to 1, namely  equal to 0.98 or 1.35 if estimated through the spectral or the LCS method, respectively. The former is very close to the recent estimate provided by \cite{Baars2020b} (0.975) where only the wall-attached eddy contribution is considered. Thus, the spectral method seems to return the correct energy rate with height for the wall-attached eddy component to the turbulence intensity, even though $\overline{uu}_{ii,A}$ can be underestimated. The values for $B_1$ are very close, namely -1.76 and -1.73 for the spectral and the LCS method, respectively. In \citep{Baars2020b}, $B_1$ for $\overline{uu}_{ii,A}^+$ was estimated equal to -2.26. 

The best fit of the experimental profiles of $\overline{uu}^+_{ii,A}$ with (\ref{eq: DiagReynStresses}) is  also performed for the vertical profiles obtained using the spectral limits proposed by \cite{Hwang2015,Hu2020}, whose results are reported in table \ref{tab. A1_B1}. The spectral boundaries of \cite{Hu2020} lead to a fitted value of $A_1$ comparable with that estimated by the authors ($0.98$ versus 1.0) while using the spectral limits proposed by \cite{Hwang2015} leads to an estimate closer to the value obtained with the LCS method, i.e. $A_1=1.22$. 

From (\ref{eq: DiagReynStresses}), we can estimate the height where $\overline{uu}_{ii,A}^+$ becomes zero, i.e. $z_{max}/\Delta_E=\exp(B_1/A_1)$, which is reported in table \ref{tab. A1_B1} and compared against the previously-reported values of $z_{max}$ estimated independently from the streamwise velocity energy spectra or the LCS. A good agreement between $z_{max}/\Delta_E$ and $\exp(B_1/A_1)$ is generally observed for the different methods used for the detection of the energy contribution associated with wall-attached eddies, meaning that the selected streamwise energy is mainly limited to wall-attached-eddy contributions.

\begin{table}
\begin{center}
\def~{\hphantom{0}}
\begin{tabular}{lcccccc}
\multicolumn{1}{l}{Reference}   & \multicolumn{2}{c}{$A_1$}        & \multicolumn{2}{c}{$B_1$}      & $e^{\left(\frac{B_1}{A_1}\right)}$ & $\dfrac{z_{max}}{\Delta_E}$  \\ 
                                & Region ($ii_A$)             & Region ($ii_B$)             & Region ($ii_A$)              & Region   ($ii_B$)    & &        \\ 
Present spectra                 & $0.98\pm0.06$      & ~~$0.53\pm0.02$    & ~~$-1.76\pm0.14$    & ~~$0.99\pm0.02$  & 0.17 & 0.17    \\
Present LCS                     & $1.35\pm0.07$      & ~~$0.30\pm0.02$    & ~~$-1.73\pm0.13$    & ~~$0.63\pm0.02$          & 0.28 &  0.31    \\
\cite{Hu2020}                   & $1.05\pm0.07$      & ~~$0.68\pm0.04$    & ~~$-0.86\pm0.10$    & ~~$1.07\pm0.03$  & 0.44 &  0.53 \\
\cite{Hwang2015}                & $1.22\pm0.06$      & ~~$0.82\pm0.04$    & ~~$-2.22\pm0.15$    & ~~$1.21\pm0.03$ & 0.16 & 0.17\\
\cite{Baars2020b}               & $0.975$            & ~~$-$              & ~~$-2.26$           & ~~$-$ & 0.10 & $-$\\
Present global                          & \multicolumn{2}{c}{$1.11\pm0.04$} & \multicolumn{2}{c}{$1.43\pm0.05$}
\end{tabular}
\caption{Values of $A_1$ and $B_1$ of (\ref{eq: DiagReynStresses}) calibrated with different methods and spectral regions. Uncertainty intervals refer to $95\%$ confidence level. \label{tab. A1_B1}}
\end{center}
\end{table}

Focusing on the energy associated with VLSMs (star markers in figure \ref{fig: Variance_AE_DE}\emph{a}), a logarithmic wall-normal trend of $\overline{uu}_{ii,B}^+$ is seemingly observed throughout the vertical range probed by the LiDAR, as already noted in \cite{Hu2020}, for both spectral and LCS methods. The results for $B_1$ and $A_1$ obtained by fitting $\overline{uu}^+_{ii,B}$ with (\ref{eq: DiagReynStresses}) are reported in table \ref{tab. A1_B1} for all the considered spectral boundaries. First, $A_1$ for the wall-detached component is significantly smaller than for the wall-attached counterpart, specifically $A_1$ equal to 0.53 and 0.3 for the spectral and LCS methods, respectively. The parameter $B_1$ is positive, while for $\overline{uu}^+_{ii,A}$ is generally negative and close to one.

It is noteworthy that the value of $\overline{uu}^+_{ii,B}$ measured by the sonic anemometer at $z/\Delta_E=0.025$ is smaller than the maximum value of $\overline{uu}^+_{ii,B}$ measured by the LiDAR at $z/\Delta_E=0.05$. This result is consistent with the work by \cite{Hu2020}, where a maximum of $\overline{uu}^+_{ii,B}$ was observed at $z/\Delta_E=0.045$. Unfortunately, more data between the minimum height probed by the LiDAR ($\approx6$ m) and the 3-m height of the sonic anemometer would be needed to draw more firm conclusions on the lower part of the vertical profile for $\overline{uu}_{ii,B}^+$.

Finally, a cumulative analysis of the streamwise turbulence intensity is reported in figure \ref{fig: Variance_AE_DE}(\emph{b}) for both spectral and LCS methods. The streamwise turbulence intensity associated with type-$\mathcal{C}$ eddies, ($\overline{uu}^+_i$ with cross symbols) is added to the energy associated with wall-attached eddies (blue and black circles), then the total streamwise turbulence intensity is achieved by adding the components associated with VLSMs.

\section{Concluding remarks}\label{sec: Conclusion}
A study of a high-Reynolds number near-neutral atmospheric surface layer (ASL) flow has been presented. The streamwise velocity was measured with a scanning Doppler pulsed wind LiDAR from a height of 6 m up to 143 m with a vertical resolution of approximately 1.08 m, and a sonic anemometer deployed at a 2-m height. The main goal of this study is to identify the energy contributions in the streamwise velocity associated with wall-attached eddies and larger structures that can be generated from their streamwise concatenations, e.g. very-large-scale motions (VLSMs) and superstructures. Furthermore, the maximum height attained by wall-attached eddies has been estimated as well.

After quality control of the LiDAR data, assessment of their statistical stationarity and convergence, quantification of the spectral-gap frequency to filter out non-turbulent motions, i.e. mesoscales, and estimation of the outer scale of turbulence, $\Delta_E$, the experimental data have mainly been interrogated through two different approaches: the analysis of the energy spectra and the linear coherence spectra (LCS) of the streamwise velocity. The main findings of the present study are summarized as follows:
\begin{enumerate}
\item The eddy classification proposed by \cite{Perry1995} (type-$\mathcal{A}$ wall-attached eddies, type-$\mathcal{B}$, e.g. VLSMs and superstructures, and type $\mathcal{C}$ small-scale Kolmogorov-like eddies), and the appearance of an inverse-power-law region in the streamwise velocity energy spectra associated with wall-attached eddies and due to the overlapping between inner-scaling and outer-scaling, has been reconciled with a micro-meteorology perspective for the classification of different regions of the streamwise velocity energy spectra \citep{Hogstrom2002}, which consist of region ($i$) following the Kolmogorov inertial scaling $k_x^{-5/3}$ ($k_x$ is the streamwise wavenumber), region ($ii$), which is denoted as the eddy surface layer, characterized by the $k_x^{-1}$ law of the energy spectra, and region ($iii$) dominated by large structures. In this work, region ($ii$) has further been decomposed into two sub-regions: a high-frequency part, ($ii_A$), dominated by wall-attached eddies and evident $k_x^{-1}$ energy spectra, and a low-frequency part, ($ii_B$), where the energy associated with VLSMs and superstructures obscures the $k_x^{-1}$ trend.

\item
Based on previous works about the LCS of the streamwise velocity induced by wall-attached eddies in turbulent boundary layers \citep[e.g.][]{Baars2017,Krug2019,Baars2020a} and the present results, an analytical model for the LCS associated with wall-attached eddies inspired by the attached eddy hypothesis (AEH) \citep{Townsend1976} has been proposed. This model applies for wall-normal positions both below and above the considered reference height and for both wall-attached eddies and VLSMs or superstructures. The model encompasses three parameters, i.e. the streamwise-wavelength/height for wall-attached eddies, i.e. their aspect ratio $\AR$, a parameter $C_1$, which represents the isolated-eddy contribution to the LCS for a given eddy population density, and an offset $C_3$. The parameter $\AR$, which is estimated to be about 14.3 from the present LiDAR data set, determines the small-wavelength boundary as a function of height of the spectral energy where contributions due to wall-attached eddies begin to build up, i.e. the spectral boundary between regions ($i$) and ($ii_A$). The parameter $C_1$ determines the spectral range over which these wall-attached-eddy contributions can be observed, i.e. $\log(\Delta \lambda_x/\Delta_E)=1/C_1$, while the maximum height attained by wall-attached eddies is $\log(z_{max}/\Delta_E)=C_3/C_1$. It is noteworthy that the estimate of $z_{max}$ from the LCS is analogous to that obtained from the AEH through the vertical law of the streamwise turbulence intensity, i.e. $\log(z_{max}/\Delta_E)=B_1/A_1$, where $A_1$ is the Townsend-Perry constant. 
For the LiDAR data set under investigation, it is found that $C_1\approx0.485$ and $C_3\approx-0.56$, which leads to a maximum height for non-null streamwise turbulence intensity associated with wall-attached eddies of $z_{max}/\Delta_E\approx0.31$. Finally, the proposed analytical LCS model enables the estimate of the spectral boundary, $\lambda_x^{th}$, between the energy associated with wall-attached eddies and that due to larger structures generated by their streamwise concatenation, i.e. VLSMs and superstructures. For the present LiDAR data set it is found $\lambda_x^{th}/\Delta_E\approx4.5$.

\item
The analysis of the streamwise velocity energy spectra has enabled us to identify the $k_x^{-1}$ region, allegedly associated with wall-attached eddies, for heights below $\approx 0.17 \Delta_E$, which is a smaller estimate than that obtained from the LCS analysis ($0.31\Delta_E$). The high-wavenumber limit, $P$, is found to follow an inertial scaling, i.e. with an  increasing wavenumber with increasing wall-normal position according to an aspect ratio of 10.8, which is smaller than the respective estimate from the LCS analysis ($\AR\approx14.3$). On the other hand, the low-wavenumber limit of region ($ii_A$), $G$, is roughly constant with height and corresponds to a wavelength of nearly $1.9\Delta_E$, again smaller than the respective LCS estimate of nearly $4.5\Delta_E$. Finally, the low-wavenumber limit of region ($ii_B$), $F$, is roughly constant with height and corresponds to a wavelength of nearly $14.3\Delta_E$. In summary, the analysis of the streamwise velocity energy spectra for the detection of the spectral regions associated with different eddy typologies seems reasonable, although encompassing a high level of uncertainty in the estimates of the actual spectral limits of the various regions and the maximum height attained by wall-attached eddies. The main sources of this uncertainty are the empirical nature of the procedure and the presence of the energy associated with larger coherent structures, such as VLSMs and superstructures, obscuring the part of the pre-multiplied energy spectra with a roughly constant energy level. Nonetheless, even though the spectral range associated with wall-attached energy and, thus, the respective turbulence intensity are underestimated, the analysis of the streamwise velocity energy spectra enables good estimates of the Townsend-Perry constant. 

\item The identification of the different regions in the streamwise velocity energy spectra through the analysis of the LCS, initially performed with a reference height $z_R=0.05\Delta_E$, has resulted to be a reliable procedure. The boundary between region ($i$) and ($ii_A$) is identified with the relationship $\lambda_x=14.3z$, namely with an aspect ratio slightly larger than that estimated through the analysis of the energy spectra ($\AR\approx10.8$), yet very close to the results reported in \cite{Baars2017} ($\AR=14$ for boundary layers). Nonetheless, the close agreement between this spectral limit identified through the LCS and the limit $P$ identified through the analysis of the energy spectra should corroborate the connection between the inverse-power-law region in the streamwise velocity energy spectra and the energy associated with wall-attached eddies. As mentioned above, the spectral limit between regions ($ii_A$) and ($ii_B$) is associated with $\lambda_x^{th}/\Delta_E\approx4.5$, which might be a better estimated than the value of 1.9 obtained from the analysis of the energy spectra, whose inverse-power-law region is obscured by the energy associated with larger coherent structures at the large-wavelength part. Finally, the LCS approach does not provide a criterion to define an analogous low-wavenumber limit of region ($ii_B$) as for the spectral analysis. The LCS approach enables the identification of the energy component incoherent with the reference height, and with $z_R/\Delta_E=0.05$, it is allegedly assumed incoherent with the wall for ASL flows. However, it is not currently possible to discern if this incoherent energy is associated with structures entrained from above, i.e. top-down mechanism, or generated at the wall and then detached, i.e. bottom-up mechanism. This limitation of the current LCS approach might be a motivation for future research. From the LCS analysis, the maximum height attained by wall-attached structure has been estimated as $z_{max}/\Delta_E\approx0.31$, which is larger than the value of 0.17 estimated through the energy spectra, yet significantly lower than the value obtained by \cite{Baars2017} for boundary layer flows ($0.7\Delta_E$). The LCS estimate of $z_{max}/\Delta_E\approx0.31$ might recall the conceptualization of an ASL flow by \cite{Hunt2000,Hogstrom2002,Drobinski2007}, who subdivided the surface layer into a lower eddy surface layer, which is dominated by wall-confined, bottom-up dynamics of coherent structures, and an upper shear surface layer, which is dominated by shear-driven, top-down motions from the mixing layer aloft. The LCS analysis would suggest that the eddy surface layer is dominated by wall-attached eddies.

\item The LCS calculated for reference heights $0.2\Delta_E\leq{z}_R\leq0.7\Delta_E$ revealed a predominance of local type-$\mathcal{C}$ eddies for heights in the proximity of $z_R$ that conceal the actual contribution of wall-attached and VLSMs \citep[consistently with the observations of][]{Krug2019,Baars2020a}. Nonetheless, for $z_R\leq z_{max}$, the LCS calculated from the LiDAR data below the reference height shows a high-wavenumber boundary between the coherent and incoherent energy components roughly invariant with height in the proximity of the the wavelength $\lambda_x=\AR z_R$, as predicted from the analytical model proposed in this paper. Furthermore, the LCS becomes gradually negligible below $z_R$ with increasing the reference height above $z_{max}$, as predicted by the AEH and the analytical LCS model. 

\item The streamwise turbulence intensity associated with wall-attached eddies, $\overline{uu}_{ii,A}$ has been assessed against the AEH prediction. The parameter $A_1$ fitted from $\overline{uu}_{ii,A}^+$ leads to estimates of 0.98 and 1.35 for the spectral and LCS methods, respectively, which are in good agreement with the recent results presented in \cite{Baars2020b}, thus confirming the scaling argument of \cite{Perry1986} for ASL flows. The scattering in the estimates of $A_1$ might be caused by the different spectral limits between regions ($ii_A$) and ($ii_B$) identified with the spectral and LSC approach, which, in turn, leads to different values of integrated energy and, thus, wall-normal distributions. On one hand, the spectral-based underestimation of this boundary ($\lambda_x=1.9\Delta_E$) leads to a downshift of the wall-normal profile of $\overline{uu}^+_{ii,A}$, yet it returns a more reliable estimate of $A_1$ (0.98) as it encompasses mainly type-$\mathcal{A}$ eddy contributions. On the other hand, the LCS-based estimate of this spectral boundary ($\lambda_x=4.5\Delta_E$) leads to a cross-influence of larger coherent structures, e.g. VLSMs and superstructure, onto the wall-normal profile of $\overline{uu}^+_{ii,A}$, and, thus, to an overestimate of $A_1$. For region ($ii_B$), a logarithmic decay of integrated energy is observed, which seemingly confirms the scenario hypothesized by \cite{Hu2020} of a geometrically similar distribution of VLSMs.
\end{enumerate}

In summarizing, this work has provided evidence that investigations of a near-neutral ASL flow with a scanning Doppler pulse wind LiDAR can open up research opportunities to investigate high Reynolds-number turbulent boundary layers, upon the optimal design of the LiDAR scanning strategy and post-processing of the generated observations. In this work, the use of the LCS has enabled the identification of the energy components either coherent or incoherent with the ground, their spectral limits with height, and the maximum height attained by wall-attached eddies. However, this current LCS approach shows symptoms of cross-contamination on the wall-attached energy contribution with those generated by type-$\mathcal{C}$ and type-$\mathcal{B}$ eddies, which might affect, for instance, the estimate of the Townsend-Perry constant. Finally, other data-driven approaches, coupled with the methods tested for this work, might provide more detailed and accurate analyses of the organization and dynamics of coherent structures in turbulent boundary layers, which might be the focus of future investigations.

\section*{Acknowledgments}
This research has been funded by the National Science Foundation, Fluid Dynamics Program, Award No. 1705837, and the NSF CAREER program, Award No. 2046160, Program Manager Ron Joslin. Michele Guala is acknowledged for providing literature data reported in figures \ref{fig: MeanVelocity} and \ref{fig: Variance}.

\appendix

\section{Evaluation of the spectral gap and outer scale of turbulence}\label{subsec: SpectralGap}
A challenge to investigating atmospheric turbulent flows is represented by the coexistence of turbulent scales of motion and background large-wavelength flow fluctuations affecting the entire boundary layer height, which are associated with the mesoscale flow component \citep[e.g.][]{Draxl2021}. Although the non-turbulent mesoscale velocity fluctuations are expected to occur with larger wavelengths than those associated with turbulence, a systematic method for mesoscale-turbulence separation is still elusive \citep{Hogstrom2002,Metzger2007}. The streamwise velocity energy spectrum typically presents a local minimum at the interface between turbulence and mesoscales, which is referred to as the “spectral gap” \citep{vanDerHoven1957,Panofsky1969,Hogstrom2002,Wyngaard2004,Metzger2007,Guala2011,Larsen2013,Larsen2016}, while the co-spectrum of the  turbulent momentum flux, $\overline{uw}$, becomes negligible for frequencies lower than the spectral-gap frequency \citep{Metzger2007}. 

\begin{figure}
    \centering
        \includegraphics[width=\textwidth]{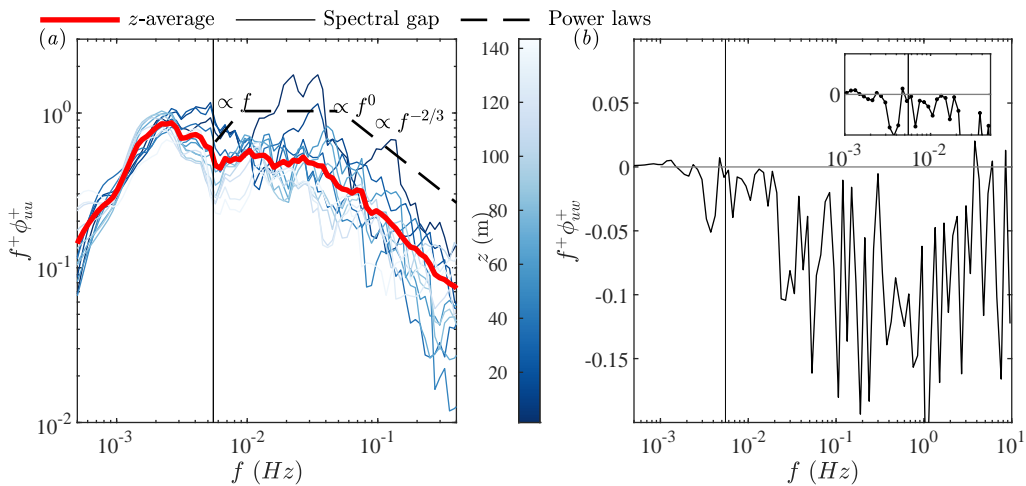}
    \caption{Detection of the spectral gap: (\textit{a}) pre-multiplied streamwise velocity energy spectra of the LiDAR data (the vertically-averaged energy spectrum is reported with a red line). (\textit{b}) co-spectrum of the vertical turbulent momentum flux measured from the “PA2” sonic anemometer. \label{fig: SpectralGap}}
\end{figure}

For this work, the pre-multiplied streamwise velocity energy spectra obtained from the wind LiDAR measurements, and the co-spectrum of the vertical turbulent momentum flux measured from the sonic anemometer “PA2” (figure \ref{fig: IPAQS}\emph{a}) are analyzed. The energy spectra reported in figure \ref{fig: SpectralGap}(\textit{a}) versus frequency, $f$, are calculated at each height sampled with the LiDAR through the Welch spectrogram \citep{Welch1967} using a window length of $0.0003$ Hz and $10\%$ overlapping between consecutive sub-periods. The energy spectra are evaluated over 100 frequencies logarithmically-spaced between 10$^{-4}$ Hz and 0.5 Hz (the Nyquist frequency), which are then smoothed through a moving-average algorithm with a spectral stencil of $f_i\pm 0.35f_i$ for a generic frequency $f_i$ \citep{Baars2020a}. The wall-normal average between all the spectra is then calculated and reported in figure \ref{fig: SpectralGap}(\textit{a}) with a red line. The pre-multiplied energy spectra of the streamwise velocity indicate the spectral gap at a frequency of about $0.0055$ Hz, which is very close to the value reported in \cite{Metzger2007} ($0.005$ Hz) for a neutrally-stratified flow probed at the SLTEST facility through a vertical array of sonic anemometers, and it is in good agreement with the averaging time of 3.3 minutes (corresponding to $0.0051$Hz) used in \cite{Guala2011} to remove mesoscale contributions to the velocity turbulence statistics.

The estimate for the spectral-gap frequency is also supported by the analysis of the co-spectrum of the vertical momentum flux measured through the PA2 sonic anemometer (figure \ref{fig: SpectralGap}\textit{b}), which is obtained through the Welch spectrogram algorithm for 500 frequencies logarithmically spaced between $10^{-4}$ Hz and 10 Hz (Nyquist value), and using a window length of $0.0003$ Hz with $10\%$ overlapping period. The co-spectrum is bin-averaged over 100 non-overlapping bins to highlight the zero-crossing region at low frequencies, which is approximately located at $0.0055$ Hz. The latter is practically equal to the spectral-gap frequency quantified through the LiDAR data. 

\begin{figure}
    \centerline{\includegraphics{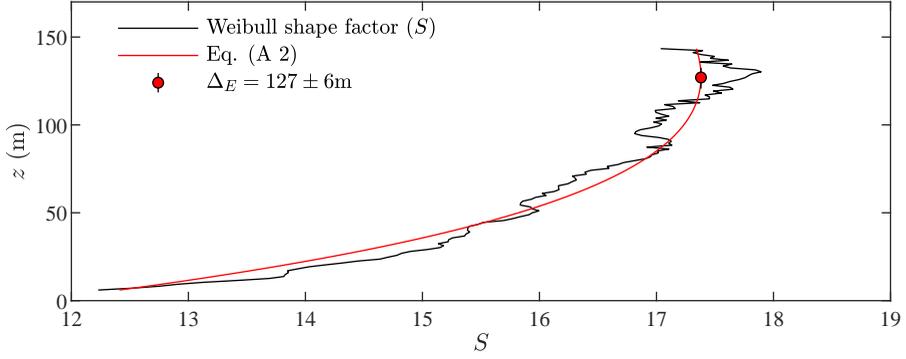}}
    \caption{Vertical profile of the Weibull shape parameter, $S$, estimated from the LiDAR velocity signals (black line) and fitted through the model of (\ref{eq: Gryning}) (red line). \label{fig:Delta_E}}
\end{figure}

The quantification of the spectral-gap frequency is also instrumental for the estimate of the outer scale of turbulence, $\Delta_E$, which is assumed as the wall-normal position where the turbulence intensity achieves a minimum value \citep{Gryning2014,Gryning2016}. For each height, the Weibull probability density function of the streamwise velocity is generated \citep{Gryning2016}: 
\begin{equation}\label{eq: Weibull}
	f_u(u) = \dfrac{S}{B}\left(\dfrac{u}{B}\right)^{S-1}~\exp\left[-\left(\dfrac{u}{B} \right)^S \right],
\end{equation}
where $S$ and $B$ represent the shape and scale parameters, respectively, and $u$ is the zero-mean velocity fluctuation. The minimum of the turbulence intensity is identified from the maximum of the shape parameter, $S$, throughout the vertical range probed by the LiDAR \citep{Gryning2016}. 
Subsequently, the vertical profile of $S$ is parametrized through the model proposed by \cite{Gryning2014,Gryning2016} for heights lower than $\Delta_E$:
\begin{equation}\label{eq: Gryning}
	S(z;~\Delta_E,c) = S_{min} + c\dfrac{z-z_{min}}{\Delta_E-z_{min}}\exp\left(-\dfrac{z-z_{min}}{\Delta_E-z_{min}} \right),
\end{equation}
where $S_{min}$ is the shape parameter associated with the lowest height probed, $z_{min}$, which is equal to 6 m for the data set under investigation. It should be noted that in \cite{Gryning2014} a second term is added to (\ref{eq: Gryning}), which refers to the shape parameter distribution above $\Delta_E$. The outer scale of turbulence and the dimensionless parameter $c$ of (\ref{eq: Gryning}) are  obtained from the least-squares fitting of the vertical profile of $S$ estimated from the LiDAR data. 
It is obtained $\Delta_E=127~\textrm{m}\pm6$ m with a 95\% confidence level, as visualized in figure \ref{fig:Delta_E}. 

\section{Spectral correction of the LiDAR velocity measurements}\label{sec: SpectraCorrection}

Wind velocity measurements performed with a Doppler wind LiDAR entail an averaging process over each measurement volume, which is mainly affected by the probe length, $l$, of each laser pulse, and the spatial distribution of the energy within the laser pulse. The radial velocity recorded at a radial distance $r$ can be modeled as the convolution between the true radial speed $\tilde{V}_r(r,t)$, namely the wind velocity component along the direction of the LiDAR laser beam, and a weighting function $\omega(r)$ representing the energy distribution along the laser pulse \citep{Frehlich1998,Frehlich2002,Mann2009,Cheynet2017,Puccioni2021}:
\begin{equation}\label{eq: LiDARconvolution}
    V_r(r,t) = \int_{-l/2}^{l/2}\tilde{V}_r(r+x',t)\omega(x')\text{d}x',
\end{equation}
For the LiDAR unit used for this experiment, $l=18$ m and $\omega=0$ outside of the range gate. The convolution in the physical domain corresponds to a product in the spectral domain between the true streamwise velocity spectrum ($\tilde{\phi}_{uu}(k_x)$) and the squared modulus of the Fourier transform of $\omega$ (said $\Omega(k_x)$) \citep{Mann2009,Puccioni2021}, such as the energy spectrum of the measured velocity signal can be estimated as:
\begin{equation}
    \phi_{uu}(k_x) = |\Omega(k_x)|^2\tilde{\phi}_{uu}(k_x).
\end{equation}
Therefore, if $|\Omega(k_x)|^2$ across frequencies was known, the low-pass filtering effect due to the LiDAR spatial averaging could be reverted to correct the streamwise velocity energy spectra and, thus, turbulence intensity. In this work, this correction is performed by using the method proposed in \cite{Puccioni2021}, which does not require any input related to the technical specifications of the used LiDAR system, such as the energy distribution over the laser pulse and LiDAR probe length, while it is completely data-driven. Specifically, the spectral model of \cite{Kaimal1972} for a neutrally-stratified ASL flow is firstly calibrated on the low-wavenumber portion of each experimental spectrum; then, the missing energy portion is quantified through the ratio between the LiDAR and the \cite{Kaimal1972} spectra and fitted with a low-pass filter model. Finally, the latter is used to revert the range-gate averaging effect onto the original spectrum.  

\begin{figure}
   \centering
    \includegraphics[width=\textwidth]{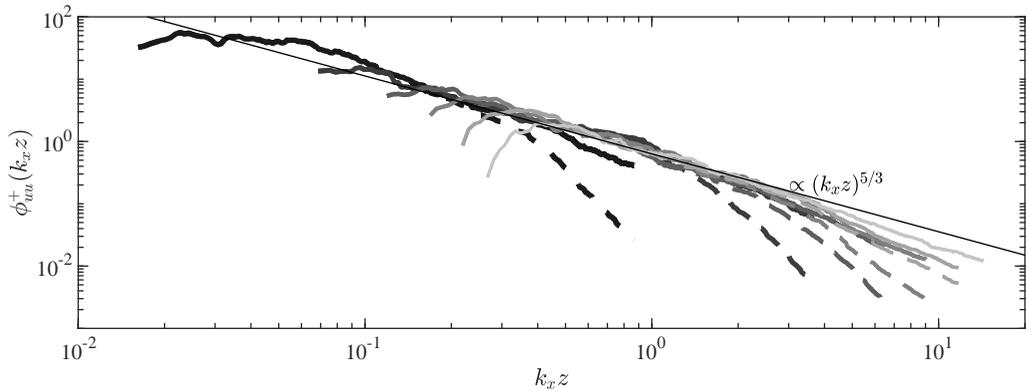}    \caption{Correction of the streamwise velocity energy spectra obtained from the LiDAR measurements at six different wall-normal positions equally spaced between 6 m and 121 m, respectively. Lines become darker with increasing height. Continuous lines represent corrected spectra, while dashed lines report raw spectra. \label{fig: SpectraCorrected}}
\end{figure}

The results of the correction procedure are reported in figure \ref{fig: SpectraCorrected} for LiDAR velocity measurements collected at six different range gates, i.e. wall-normal positions. It is observed that, after the spectral correction, the expected $k_x^{-5/3}$ slope is roughly recovered for all the velocity signals, together with a good overlapping on the same power-law when the spectra are reported as a function of the inertia-scaled wavenumber, $k_x z$ \citep{Perry1986}.

For the lowest part of the ASL ($z<l$,), it should be noticed that the present correction method still underestimates the spectral energy, as the $k_x^{-1}$ region is also affected by the LiDAR spatial averaging. Other spectral correction procedures were tested \citep{Mann2009,Cheynet2017}, while the data-driven method of \cite{Puccioni2021} provided the largest variance recovery. 

\section{Smoothing of the linear coherence spectrum}\label{app: Smoothing}
\begin{figure}
    \centerline{\includegraphics{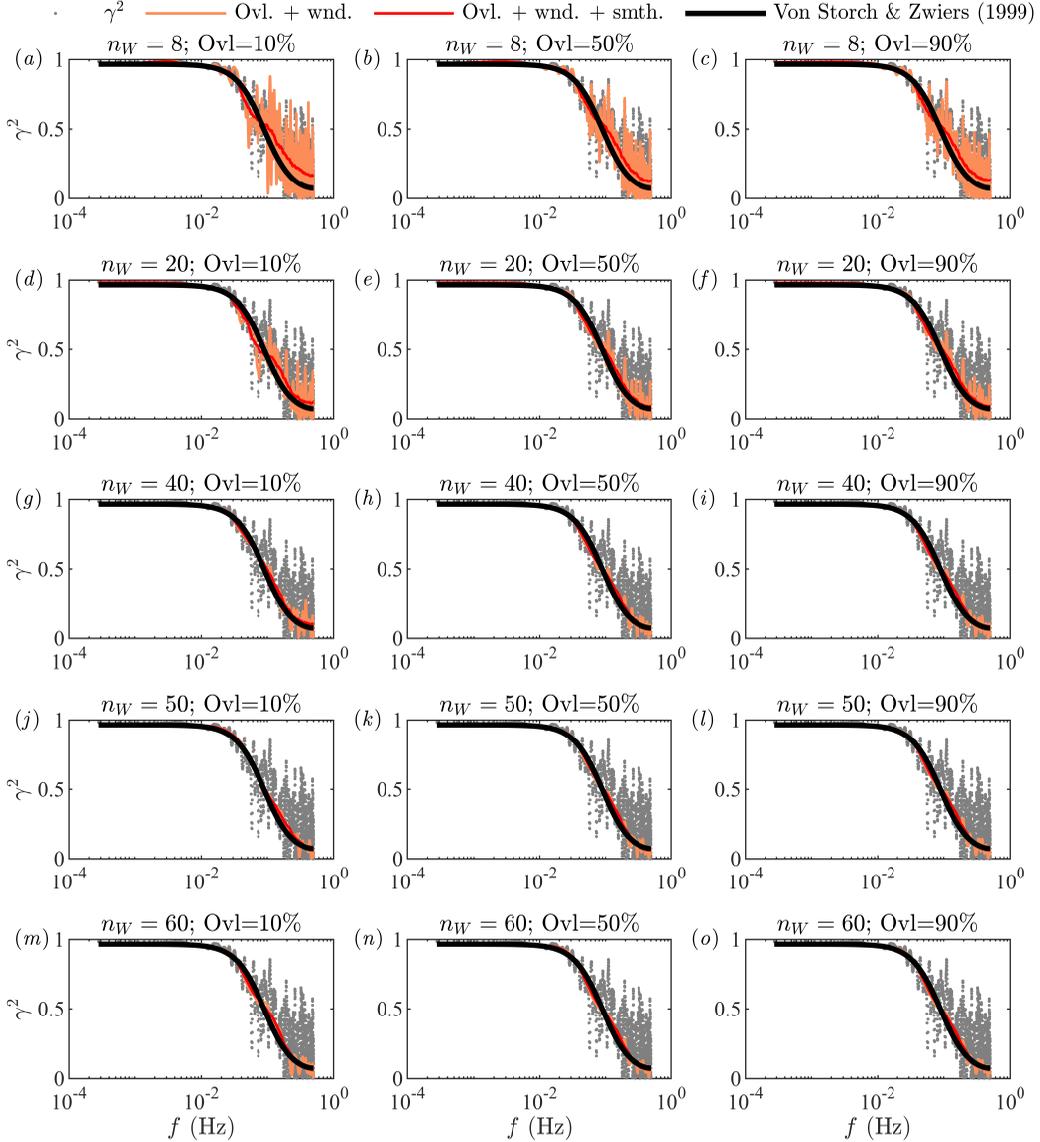}}
    \caption{Sensitivity of the linear coherence spectrogram (LCS) to the windowing and overlapping parameters for a synthetic autoregressive case. Grey dots report the LCS without smoothing, the orange lines depict $\gamma^2$ calculated with the \cite{Welch1967} periodogram with the indicated windowing and smoothing parameters, the red lines reports the result after the smoothing procedure, and the black lines are the theoretical distribution of \cite{VonStorch1999}.}\label{fig: LCS_all}
\end{figure}

The parameters for the Welch spectral estimator \citep{Welch1967}, namely window length and overlapping, used for the calculation of the linear coherence spectrum, LCS, are firstly determined by leveraging a synthetic first-order autoregressive model, $q_n$, defined as follows \citep{VonStorch1999}:
\begin{equation}\label{eq: AR1}
    q_n = \left\{\begin{array}{ll}
         e_n&~\text{as}~n=1  \\
         \alpha q_{n-1} + e_n&~\text{as}~n>1 
    \end{array}
    \right.,
\end{equation}
where $e_n$ is a white-noise Gaussian process with variance $\sigma_e^2$,  $0<\alpha<1$ and $n=1,...,M$. From (\ref{eq: AR1}) a second synthetic signal can be defined as:
\begin{equation}\label{eq: AR1_y}
    s_n = \beta q_n + \tilde{e}_{n},
\end{equation}
where $\tilde{e}_{n}$ is another white-noise Gaussian process with variance $\tilde{\sigma}_{e}^2$, and $0<\beta<1$. From the definition of auto- and cross-spectra, it is possible to show that the coherence between $q$ and $s$ is given by:
\begin{equation}\label{eq: vonStorch}
    \gamma^2(q,s;f)=\dfrac{\beta^2\phi_{qq}(f)}{\beta^2\phi_{qq}(f)+\tilde{\sigma}_{e}^2},
\end{equation}
where $f$ is the frequency and:
\begin{equation}
    \phi_{qq}(f) = \dfrac{\sigma_e^2}{1+\alpha^2-2\alpha\cos(2\pi f)}.
\end{equation}
Thus, leveraging these two related autoregressive signals $q_n$ and $s_n$, a theoretical reference for the LCS is available against which the numerical algorithm for the evaluation of the LCS from the two signals can be assessed. 

As previously mentioned, the coherence for two time series is calculated by means of the \cite{Welch1967} algorithm; the ensemble averaging operation in (\ref{eq: LCS}) is simulated by dividing the entire signals in a certain number of sub-windows, $n_W$, with a certain overlapping percentage. Successively, a moving average with stencil $f_n\pm0.35f_n~(n=1,...,M)$ is performed to smooth the LCS \citep{Baars2017,Baars2020a}. 

The algorithm for the calculation of the LCS between the discrete signals of (\ref{eq: AR1}) and (\ref{eq: AR1_y}) is tested, then the results are compared against the analytical LCS reported in (\ref{eq: vonStorch}). The results obtained from this sensitivity study are reported in figure \ref{fig: LCS_all} using: $\alpha=0.9;~\beta=0.3$ for a signal with $M=3600$ samples simulated with a sampling frequency of $1$ Hz. A close agreement between the numerical results and the analytical prediction is obtained for the case with $n_W=20$ and $90\%$ overlapping (figure \ref{fig: LCS_all}\emph{f}), which are the parameters used for the present work. 

\bibliographystyle{jfm}
\bibliography{Bibliography}
\end{document}